\documentclass[12pt]{iopart}
\usepackage{graphicx}
\expandafter\let\csname equation*\endcsname\relax
\expandafter\let\csname endequation*\endcsname\relax
\usepackage{amsmath}
\usepackage{mathabx}
\usepackage{cite}
\usepackage{lscape}
\usepackage{varwidth}
\usepackage{textcomp}
\usepackage{xcolor,soul}

\begin{document}

\title[MICROSCOPE first results]{Space test of the Equivalence Principle: first results of the MICROSCOPE mission}

\author{Pierre Touboul$^1$, Gilles M\'etris$^2$, Manuel Rodrigues$^1$, Yves Andr\'e$^3$, Quentin Baghi$^2$\footnote{Current address: NASA Goddard Space Flight Center, Greenbelt, MD 20771, USA}, Joel Berg\'e$^1$, Damien Boulanger$^1$, Stefanie Bremer$^4$\footnote{Current adress: DLR, Institute of Space Systems, Robert-Hooke-Str. 7, 28359 Bremen}, Ratana Chhun$^1$, Bruno Christophe$^1$, Valerio Cipolla$^3$, Thibault Damour$^5$, Pascale Danto$^3$, Hansjoerg Dittus$^6$, Pierre Fayet$^7$, Bernard Foulon$^1$, Pierre-Yves Guidotti$^3$, Emilie Hardy$^1$, Phuong-Anh Huynh$^1$, Claus L\"ammerzahl$^4$, Vincent Lebat$^1$, Fran\c{c}oise Liorzou$^1$, Meike List$^4$, Isabelle Panet$^8$, Sandrine Pires$^{9}$, Benjamin Pouilloux$^3$, Pascal Prieur$^3$, Serge Reynaud$^{10}$, Benny Rievers$^4$, Alain Robert$^3$, Hanns Selig$^4$\footnote{Current adress: GERADTS GMBH, Kleiner Ort 8, D-28357 Bremen, Germany}, Laura Serron$^2$, Timothy Sumner$^{11}$, Pieter Visser$^{12}$}

\address{$^1$ DPHY, ONERA, Universit\'e Paris Saclay, F-92322 Ch\^atillon, France}
\address{$^2$ Universit\'e C\^ote d{'}Azur, Observatoire de la C\^ote d'Azur, CNRS, IRD, G\'eoazur, 250 avenue Albert Einstein, F-06560 Valbonne, France}
\address{$^3$ CNES, 18 avenue E Belin, F-31401 Toulouse, France}
\address{$^4$ ZARM, Center of Applied Space Technology and Microgravity, University of Bremen, Am Fallturm, D-28359 Bremen, Germany}
\address{$^5$ IHES, Institut des Hautes Etudes Scientifiques, 35 Route de Chartres, 91440 Bures-sur-Yvette, France}
\address{$^6$ DLR, K\"oln headquarters, Linder H\"ohe, 51147 K\"oln, Germany}
\address{$^7$ LPTENS, Ecole Normale Sup\'erieure (PSL, CNRS, SU, UPD-USPC, Paris, France), and CPhT, Ecole Polytechnique (Palaiseau, France)}
\address{$^8$ IGN, Institut g\'eographique national, 73 avenue de Paris, F-94160 Saint Mand\'e, France}
\address{$^{9}$ Laboratoire AIM, CEA, CNRS, Universit\'e Paris Saclay,  Universit\'e Paris Diderot, Sorbonne Paris Cit\'e, F-91191, Gif-sur-Yvette, France}
\address{$^{10}$ Laboratoire Kastler Brossel, UPMC-Sorbonne Universit\'e, CNRS, ENS-PSL University, Coll\`ege de France, 75252 Paris, France}
\address{$^{11}$ Blackett Laboratory, Imperial College London, Prince Consort Road, London. SW7 2AZ, United Kingdom}
\address{$^{12}$ Faculty of Aerospace Engineering, Delft University of Technology, Kluyverweg 1, 2629 HS Delft, Netherlands} 
\ead{pierre.touboul@onera.fr, gilles.metris@oca.eu, manuel.rodrigues@onera.fr}
\vspace{10pt}
\begin{indented}
\item[] Jan. 2019
\end{indented}

\begin{abstract}
The Weak Equivalence Principle (WEP), stating that two bodies of different compositions and/or mass fall at the same rate in a gravitational field (universality of free fall), is at the very foundation of General Relativity. The MICROSCOPE mission aims to test its validity to a precision of $10^{-15}$, two orders of magnitude better than current on-ground tests, by using two masses of different compositions (titanium and platinum alloys) on a quasi-circular trajectory around the Earth. This is realised by measuring the accelerations inferred from the forces required to maintain the two masses exactly in the same orbit.
Any significant difference between the measured accelerations, occurring at a defined frequency, would correspond to the detection of a violation of the WEP, or to the discovery of a tiny new type of force added to gravity. MICROSCOPE's first results show no hint for such a difference, expressed in terms of E\"otv\"os parameter $\delta(Ti,Pt)=[-1\pm{}9{\rm (stat)}\pm{}9{\rm (syst)}] \times{}10^{-15}$ (both 1$\sigma$ uncertainties) for a titanium and platinum pair of materials. This result was obtained on a session with 120 orbital revolutions representing 7\% of the current available data acquired during the whole mission. The quadratic combination of 1$\sigma$ uncertainties leads to a current limit on $\delta$ of about $1.3\times{}10^{-14}$.
\end{abstract}


\submitto{\CQG}
%
%
%

\section{Introduction}

A hundred years ago, Einstein's theory of General Relativity (GR) \cite{einstein08, einstein16} revolutionised our understanding of gravitation, transforming the well-known ``at-distance" force into a manifestation of the interplay between matter and the curved space-time manifold. The newborn theory was eagerly accepted after it solved the Mercury perihelion puzzle and Eddington measured the gravitational deflection of stars' light passing near the Sun. But its most exotic predictions were the existence of gravitational waves and of black holes. The former were indirectly discovered from the observed decrease in the period of the Hulse-Taylor pulsar \cite{hulse75} in the 1970s, before LIGO's direct detection in 2015 \cite{abbott16}; the gravitational waves observed during this event were produced by the merger of two black holes, thereby proving the existence of the latter. Today, GR has passed all experimental tests and seems unassailable.
A few years before, CERN's Large Hadron Collider had found the last missing piece in the Standard Model, the Brout-Englert-Higgs boson, with a mass of 125 GeV \cite{atlas12, cms12}.

However, despite those successes, it is hardly the end of the route for fundamental physics. Shedding light on the dark sector is proving particularly difficult, decades after the discovery of the missing mass at cosmological scale \cite{zwicky33, rubin70} and of the acceleration of the cosmic expansion \cite{riess98, perlmutter99}. Other questions remain unanswered, dealing in particular with symmetries and symmetry-breaking, the possibility of a supersymmetry between bosons and fermions through a supersymmetric extension of the Standard Model \cite{fayet77}, the origin of the preponderance of matter over antimatter, or the problems of quantum gravity and the quest for a possible unification of all interactions.

Theories beyond the standard model propose the existence of new particles. For instance, string-inspired theories introduce a spin-0 dilaton-like particle (e.g. Refs. \cite{damour94, damour02}), and extensions of the Standard Model gauge group suggest the possible existence of a very light spin-1 U-boson mediating a new force \cite{fayet90, fayet17}. Other models, such as scalar-tensor models, modify GR's equations via the introduction of a new scalar field (see e.g. Refs. \cite{damour92, clifton12, joyce15}). The existence of a new very light scalar field (thereby, of a new long-range force) can be made compatible with current solar system tests with the inclusion of a screening mechanism that makes the field's mass environment-dependent \cite{damour94,vainshtein72,Damour:1992kf,khoury04a, khoury04b, babichev09,hinterbichler10, brax13,burrage18}. Although they mimic GR because of their screening mechanism, those models can nevertheless have measurable effects, such as an apparent violation of the equivalence principle (e.g. \cite{khoury04b, damour12}).

The weak equivalence principle (WEP) states that two bodies of different compositions and/or masses fall at the same rate in the same gravitational field (universality of free fall-UFF); similarly, it states the equivalence of the ``inertial'' and ``gravitational'' masses. It was formulated by Einstein in 1907 as a starting point of GR, and has since been verified experimentally with higher and higher precision. Tests of the WEP are usually presented in terms of the E\"otv\"os ratio $\eta$ \cite{eotvos22}, defined as the normalised difference of acceleration (or equivalently, as the normalised difference of gravitational-to-inertial masses) of two test bodies affected by the same gravitational field \cite{will14}:
\begin{equation} \label{eq_eotvos}
\eta = 2 \frac{a_2-a_1}{a_2+a_1} = 2 \frac{m_{g2}/m_{i2} - m_{g1}/m_{i1}}{m_{g2}/m_{i2} + m_{g1}/m_{i1}}
\end{equation}
where $a_j$ is the acceleration of the $j$th test-body, and $m_{g,j}$ and $m_{i,j}$ are its gravitational and inertial masses. In this paper, we use a good first order approximation of the E\"otv\"os parameter $\delta (2,1)$:
\begin{equation}
\delta(2,1) \equiv \frac{m_{g2}}{m_{i2}} - \frac{m_{g1}}{m_{i1}}.
\end{equation}

Tests of the UFF and of the WEP have a long history, starting with Galileo Galileo (1638) and Newton (1687), and continuing to the end of the 20th century after Fischbach \cite{fischbach86} revived the interest in experimental searches for new, WEP-violating interactions.
The state-of-the-art experiments have measured $|\eta|<{\rm a \,\, few\,\,} 10^{-13}$ (see Ref. \cite{will14} for a historical account of tests of the WEP): (i) the E\"ot-Wash group used a high-precision torsion pendulum in the Earth and Sun gravitational fields \cite{schlamminger08, wagner12}, and (ii) the Lunar Laser Ranging has monitored the motion of the Moon and the Earth around the Sun \cite{williams12,viswanathan18} and measures a combination of the WEP and SEP with a slightly better accuracy.

However, in spite of huge efforts to incrementally improve these experiments, it became apparent in the early 2000's that a new approach was needed to significantly improve on existing constraints on the WEP. In the 1970's, Chapman \cite{chapman01} proposed a space experiment to test the Equivalence Principle. It was the basis of the STEP experiment extensively studied in Stanford University \cite{everitt2003}. Performing a test in space became feasible with ultra-sensitive accelerometers and drag-free satellites, as experimented with GRACE \cite{tapley04}, GOCE \cite{rummel11} and LISA Pathfinder \cite{armano16}. Thence, those technologies were shown to be well suited to measure weak accelerations in a well-controlled dynamical environment motion \cite{touboul04}.

In the early 2000s, the Centre National d'Etudes Spatiales (CNES), the Observatoire de la C\^ote d{'}Azur (OCA) and the Office National d'Etudes et de Recherches A\'erospatiales (ONERA) embarked on the development of the MICROSCOPE (Micro-Satellite \`a tra\^in\'ee Compens\'ee pour l'Observation du Principe d'Equivalence) mission \cite{touboul99, touboul01a, touboul01b,touboul09,touboul12}, the first laboratory experiment that would actually test the WEP in space.
The experiment relies on the comparison of the free-fall motion of two test-masses of different composition (one of titanium alloy and one of platinum alloy) at the centre of a dedicated drag-free and attitude-controlled satellite (see Sect. \ref{sect:ssect_drag}). At the core of the instrument, an ultra-sensitive accelerometer forces each test mass to remain in equilibrium using electrostatic forces. Thus, the test-masses have to follow the motion of the satellite and the electrostatic forces compensate the difference of acceleration between the masses and the satellite. In this paper, we define the electrostatic acceleration (or sometime acceleration for short) as the electrostatic force divided by the mass even if the mass are motionless with respect to the satellite.

Once potential disturbing effects are accounted for, the comparison of those electrostatic forces is a direct measure of the difference in the control accelerations of the test masses, and  hence of the WEP. If the WEP is violated, since the gravitational source is the Earth, then the measured difference will be modulated by the motion and attitude of the spacecraft along its orbit. Therefore, the violation signal will be detectable at a given frequency $f_{\rm EP}$ that is the sum of the orbital frequency and of the satellite spinning frequency. 

MICROSCOPE was launched into a low-Earth sun-synchronous orbit from Kourou on April 25, 2016 at an altitude of 710 km. The science experiment started in December 2016 after a successful commissioning phase \cite{prieur17,berge17a}. Since then, MICROSCOPE has delivered high-quality data.
In Ref. \cite{touboul17}, we used 7\% of the total data expected from the mission to provide first, intermediate results. We found no violation of the WEP, but even this small amount of data allowed us to improve the constraints on $\delta$ by one order of magnitude, down to 
\begin{equation}
\delta({\rm Ti,Pt})=[-1\pm{}9{\rm (stat)}\pm{}9{\rm (syst)}] \times{}10^{-15}
\end{equation}
at 1$\sigma$ statistical uncertainty, for the titanium and platinum pair of materials. This new upper bound on the WEP has allowed new limits to be set on beyond-GR models involving a light dilaton \cite{berge18} or a U-boson \cite{fayet18,fayet19}.

This paper is an expanded version of the letter \cite{touboul17}. We provide more details on the experiment, the instrument geometry and electronic characterisation, the assessment of systematic uncertainties and the data analysis. 
Several upcoming papers are in preparation to better detail the mission rationale and all the main subsystems relevant to the final performance. 

The layout of the paper is as follows. Sect. \ref{sect_instrument} presents the satellite and instrument, whose characteristics were assessed during the commissioning phase, as shown in Sect. \ref{sect_commissioning}. The measurement principle and systematic errors are discussed in Sect. \ref{sect_measure}. {Sect. \ref{sect_analysis} presents the data analysis (restricted to one measurement session) and discusses MICROSCOPE's first results. We conclude in Sect. \ref{sect_ccl}.}


\section{Experimental apparatus} \label{sect_instrument}

The MICROSCOPE satellite carries the T-SAGE (Twin Space Accelerometers for Gravitation Experiment) science payload, a pair of double electrostatic accelerometers  designed to test the WEP in space. In this section, we first briefly present the satellite. We then give detailed metrology and electronic information about the T-SAGE instrument, as measured on the ground before MICROSCOPE's launch. The left panel of Fig. \ref{fig_satellite} shows the satellite during its pre-launch tests; the right panel shows T-SAGE, which sits at the centre of the satellite.

\subsection{Satellite and its acceleration and attitude control system} \label{sect:ssect_drag}

\begin{figure}
\begin{center}
\includegraphics[width=0.56\textwidth]{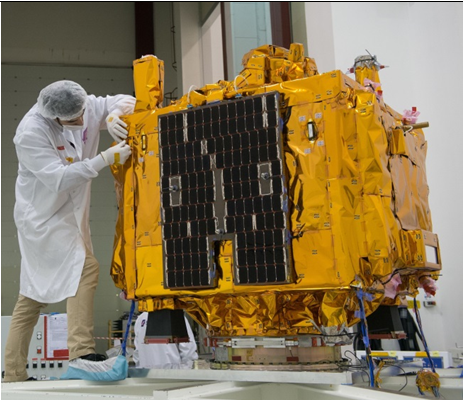}
\includegraphics[width=0.4\textwidth]{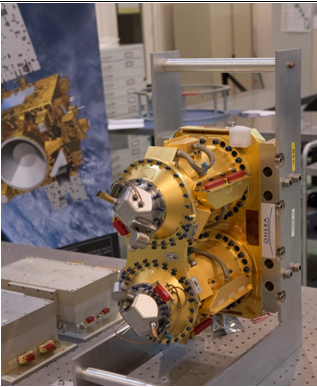}
\caption{Left: MICROSCOPE satellite during vibration test campaign (\textcopyright CNES GRIMAULT Emmanuel 2015). Right: T-SAGE payload sensor units and front end electronics in satellite clean room before integration (\textcopyright CNES/S. Girard, 2014).}
\label{fig_satellite}       
\end{center}
\end{figure}

The satellite is based on the CNES Myriade line, with a mass of 300kg and a volume of 2m$^3$ (Fig.\ref{fig_satellite}). It is covered by Multi Layer Insulation (MLI) that provides good radiative thermal filtering. Cold gas proportional thrusters are used to reduce non-gravitational accelerations experienced by the satellite and to finely control the attitude. The Drag Free and Attitude Control System (DFACS) uses the scientific instrument itself in a control loop for sensing linear and angular accelerations \cite{jafry01, theil01, prieur17}. The DFACS cancels the linear common mode acceleration in the frequency band of interest which could be measured differently by each test-mass due to the different transfer functions. 

\subsection{T-SAGE instrument: test-masses' metrology and servo-loop electronics}

\subsubsection{Sensor units}

The science payload comprises two sensor units (two SU) shown in right panel of Fig. \ref{fig_satellite}.
Both SU share the same design (sensor mechanics and electronic circuits), the same technologies (mechanics and components) and the same materials. Their only difference is in the composition of their test-masses. 

\begin{figure}
\includegraphics[width=0.95\textwidth]{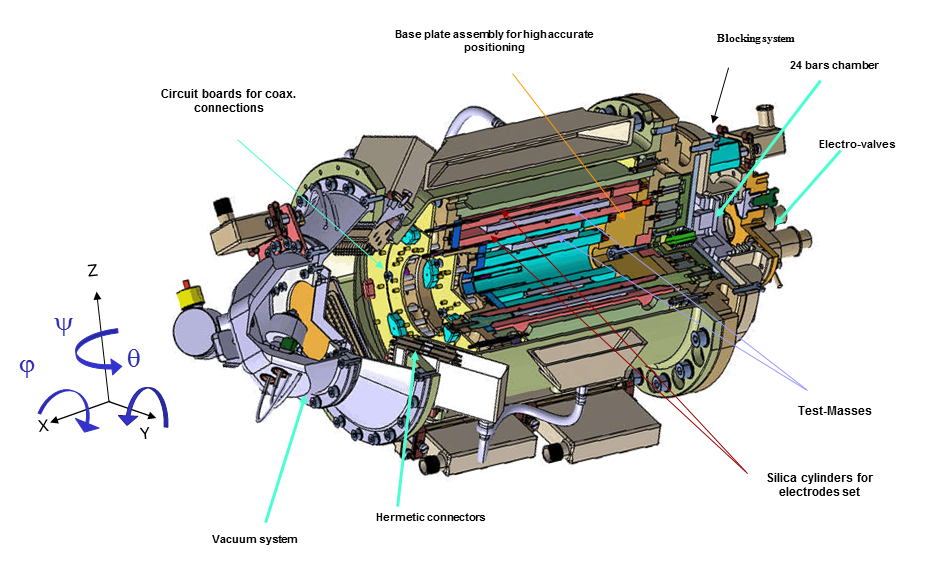}
\caption{Cut-away representation of the sensor unit. Each test-mass has its measurement reference frame symbolised by the 6 axis schematic.}
\label{fig_SU}       
\end{figure}

Fig. \ref{fig_SU} shows a cut-away view of a SU. It contains two concentric cylindrical accelerometers. 
Each accelerometer uses electrostatic levitation of a cylindrical test mass (purple cylinders in Fig. \ref{fig_SU}): pairs of electrodes (supported by silica cylinders --in red in Fig. \ref{fig_SU}) surrounding the mass and controlling the electric field arround it. The electric field generates electrostatic (negative) pressures on the test-mass, whose six degrees of freedom are digitally controlled by six independent servo-channels using different combinations of electrode pairs.
A thin gold wire of $7\rm\mu$m diameter and of $\sim$ 25mm length is glued onto each test-mass: it allows the test-mass charge control and the capacitive sensing through the application of a DC and a 100kHz voltages.

Two Front End Electronics Unit (FEEU) boxes (one per sensor unit) include the capacitive sensing of the test-mass motion, the reference voltage sources and the analog electronics to generate the voltages applied to the electrodes. An Interface Control Unit (ICU) includes the digital electronics associated with the servo-loop digital control laws, as well as the interfaces to the satellite's data bus. The FEEU output is used by the DFACS.

One sensor unit (SUREF) serves as a reference for the experiment. Its test-masses are made of the same platinum alloy (see below), so that it should not be affected by composition-dependent forces. Although not a direct probe of systematic uncertainties, it provides valuable indications about instrumental effects. In the remainder of this paper, SUREF's inner (outer) mass is called IS1-SUREF (IS2-SUREF).
The second sensor unit (SUEP) has two masses of different compositions and is used for the Equivalence Principle test. Its inner mass is made of the same platinum alloy as SUREF's test-masses, while its outer test-mass is made of titanium alloy. In the remainder of this paper, SUEP's inner (outer) mass is called IS1-SUEP (IS2-SUEP).

Each sensor unit comprises a hermetic Invar housing surrounding the silica core which is maintained under vacuum by a getter material in orbit. On ground, an ion pump is used during all the flight configuration test phases except thermal, vibration and shock qualification.

Finally, Fig. \ref{fig_SU} shows the coordinate system used in the instrument: $X$ is the main axis of the cylinder. It is the most sensitive axis (see below) and the WEP signal is estimated along this axis. Measurements along the $X$, $Y$ and $Z$ axes are used by the DFACS.

\subsubsection{Test-masses}

The cylindrical test-masses are what differentiates the sensor units. They have been produced and precisely characterised in the Physikalisch-Technische Bundesanstalt (PTB) laboratory, in Braunschweig, the German National Metrology Institute, with a metrology accuracy better than 1$\mu$m \cite{hagedorn13}.
The SU PtRh10 platinum-rhodium alloy contains 90\% by mass of Pt (A=195.1, Z=78) and 10\% Rh (A=102.9, Z=45). The isotopic composition of Pt has been measured by PTB on a sample of flight material (see Table \ref{tab_Ptiso}).
SUEP's outer test-mass is made of 90\% of titanium (A=47.9, Z=22), 6\% of aluminium (A=27.0, Z=13) and 4\% of vanadium (A=50.9, Z=23). The choice of the materials is a trade-off between the machining laboratory know-how and the theoretical motivation \cite{touboul01b, blaser01}. Titanium and platinum differ mainly from the neutron excess over the atomic mass $(N-Z)/A$ and a little from the nuclear electrostatic energy $Z(Z-1)/(N+Z)^{1/3}$.

\begin{table}
\caption{\label{tab_Ptiso} Measured isotopic composition of Pt in PtRh10 material. }
\begin{indented}
\item[]\begin{tabular}{@{}lc}
\br
Isotope & Mol per mol of PtRh10  \\
\mr
 Pt(190) & 0.000117  \\
 Pt(192) &	0.00782 \\
Pt(194) &	0.32863 \\
Pt(195) &	0.33776 \\
Pt(196) &	0.25210 \\
Pt(198) &	0.07357 \\
\br
\end{tabular}
\end{indented}
\end{table}

All test-masses have four small flat areas along their $X$-axis to break the cylindrical symmetry and to allow for angular control about $X$ ($\Phi$ angle). Their length has been optimised to keep quasi-identical moments of inertia about their three axes. The moments of inertia have been computed taking into account the measured dimensions and densities, and their dispersions. The total relative dispersion of the moments, with respect to an ideal homogenous spherical test-mass, is, in the worst case, $10^{-3}$. This is small enough to mitigate the effect of local gravity gradients as required.

The mass of each test-mass was measured with a maximum error of 0.025 mg. Density was estimated to better than 0.001g/cm$^3$ by cutting two slices from either end of each test-mass (also allowing estimation of the material homogeneity).

Table \ref{tab_metrology} summarises the test-mass metrology data. The accuracy of production of individual parts, and subsequent integration are at the micro-meter level. In particular the relative positions of the test-masses have been evaluated by direct metrology and capacitive measurements during integration. 

Finally, in order to limit the residual effect of the Earth's gravity gradient variations at the WEP test frequency, $f_{\rm EP}$, the relative centring of the test-masses was specified to be $<\,20\,\mu$m along each axis for each sensor. In Sect. \ref{sect_measure}, we show that we are able to estimate the off-centring in flight to better than 0.1$\mu$m.

\begin{table}
\caption{\label{tab_metrology} Main test-masses physical parameters measured in the laboratory before flight.}
\begin{indented}
\item[]\begin{tabular}{@{}lllll}
\br
Parameter & IS1-SUREF & IS2-SUREF & IS1-SUEP & IS2-SUEP \\
\mr
Inner radius [mm] & 30.801 & 60.799 & 30.801 & 60.802 \\
Outer radius [mm] & 39.390 & 69.397 & 39.390 & 69.401 \\
Length [mm] & 43.331 & 79.821 & 43.330 & 79.831 \\
Inertia about X [kg\,mm$^2$] & 125.0206 & 1442.454 & 125.0775 & 319.0266 \\
Inertia about Y [kg\,mm$^2$] & 125.0021 & 1442.139 & 125.0524 & 318.9978 \\
Inertia about Z [kg\,mm$^2$] & 125.0070 & 1442.214 & 125.0549 & 318.9867 \\
Maximum relative difference & 0.0004 & 0.0007 & 0.001 & 0.0001 \\
in moment of inertia \\
Mass [kg] & 0.401533 & 1.359813 & 0.401706 & 0.300939 \\
Density @ 20$^{\rm o}$C [g\,cm$^{-3}$] & 19.967 & 19.980 & 19.972 & 4.420 \\
Density homogeneity along X & 0.04\% & 0.05\% & 0.1\% & 0.001\% \\
\br
\end{tabular}
\end{indented}
\end{table}

\subsubsection{Capacitive Sensing and Electronic control}

\begin{figure}
\begin{center}
\includegraphics[width=0.75\textwidth]{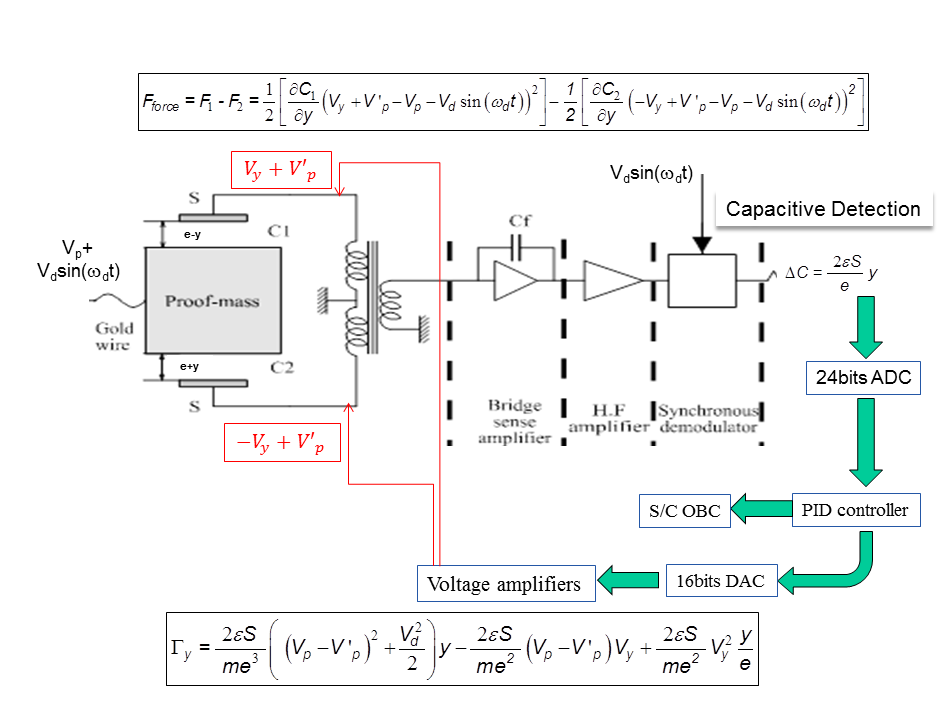}
\caption{Schematic of one degree of freedom servo-loop control.}
\label{fig_detect}       
\end{center}
\end{figure}

Each test-mass is equipped with electronics to control its movements. 
Each servo-channel is composed of (Fig.\ref{fig_detect}):
\begin{itemize}
\item capacitive sensors which measure six test-mass degrees of freedom:  three positions ($x,y,z$) and  three angles ($\phi, \theta, \psi$) about those axes \cite{josselin99}
\item a digital PID controller, whose control laws are programmed into Digital Signal Processor (DSP); the DSP has a 20 MHz cycle and operates the servo-loop at a submultiple of this frequency (1027 Hz); it computes signals representative of the forces and torques applied to the test mass and delivers them to the satellite on-board computer
\item actuators which apply voltages onto each electrode to generate the required force or couple; these voltages are obtained by digital to analogue conversion of the DSP outputs.
\end{itemize}

The inertial sensor acceleration range is limited by the voltages that can be applied to the electrodes, and it must exceed the weak residual accelerations managed by the satellite control. The voltage applied on the electrodes depends on the geometrical and electrical configuration and on the mass of the test mass; it can be expressed as an  acceleration resulting from an electrostatic force:
\begin{equation}
\Gamma_{\rm elec} = \alpha (V_p - V'_p)V_e,
\end{equation}
where $V_p$ is the DC voltage applied directly to the test mass, $V'_p$ is the offset voltage applied to the electrodes, $V_e$ is the controllable part of the voltage that can be applied to the electrodes ($\lvert V_e \rvert<40$V) and $\alpha$ is an electrostatic physical gain expressed in $\rm ms^{-2}V^{-2}$.

When  $V'_p$ is null, the voltage $V_p$ can modify the electrostatic acceleration's behaviour: the higher $V_p$, the easier the acquisition of the test-mass. For example, in order to enable the test-mass acquisition in ZARM's Bremen Tower, the $V_p$ voltage and the maximum electrode range voltage $V_{\rm max}$ were both fixed to 90V using the Engineering Model Electronics \cite{selig10}. For the Flight Electronics, $V_{\rm max}$ is limited to 40V for derating reasons. Hence, the flight model range was not compatible with the free-fall residual accelerations in the short time of fall in the ZARM tower (9 seconds in the catapult mode). 
For flight, two modes were defined : the Full Range Mode (FRM) has a relatively high $V_p=40$V and is used to acquire the test-mass after it is unlocked or the control is lost while the High Resolution Mode (HRM) uses $V_p=5$V and is used for fine science measurements. In all cases $V_{\rm max}=40$V. 
 
$V'_p$  is used to increase or decrease, depending on its sign, the scale factor for each axis.
When the electrostatic servo-loop operates properly and thus the test-mass is motionless, the electrostatic acceleration range is proportional to ($V_p-V'_p)V_{\rm max}$ to first order. Depending on the axis, $V'_p$ and $V_{\rm max}$ can be set in order to optimise the resolution versus the range.
Table \ref{tab_range} gives the $V'_p$ values and ranges for the HRM used during scientific sessions:  the differences between SUREF's inner and outer masses are due to their different size;  the differences between SUEP's and SUREF's outer masses arise from their different mass.
In HRM, $V_p$ is measured in flight as $V_p=5.003 \pm{}0.013$V with respect to the FEEU null voltage reference point (which is different from the electrical ground of the structure). The DC voltage $V'_p$ is applied symmetrically on each electrode pair. 

\begin{table}
\caption{\label{tab_range} DC $V'_p$ voltage applied symmetrically on each electrode pair and full-scale range of the applied electrostatic accelerations of the inertial sensor for each axis, in High Resolution Mode (HRM).}
\begin{indented}
\item[]\begin{tabular}{@{}lcccc}
\br
& IS1-SUREF & IS2-SUREF & IS1-SUEP & IS2-SUEP \\
\mr
$V'_p$ [V] \\
$X$ & -5 & -10 & -5 & 0 \\
$Y$ and $\Psi$ & 2.5 & 0 & 2.5 & 2.5 \\
$Z$ and $\Theta$ & 2.5 & 0 & 2.5 & 2.5 \\
$\Phi$ & -10 & -10 & -10 & -10 \\ 
\mr
Range \\
$X$ [$\mu$m\,s$^{-2}$] & 2.40 & 1.60 & 2.40 & 3.20 \\
$Y$ [$\mu$m\,s$^{-2}$] & 3.44 & 10.5 & 3.44 & 21.2 \\
$X$ [$\mu$m\,s$^{-2}$] & 3.44 & 10.5 & 3.44 & 21.2 \\
$\Phi$ [$\mu$rad\,s$^{-2}$] & 62.3 & 38.4 & 62.4 & 173 \\
$\Theta$ [$\mu$rad\,s$^{-2}$] & 62.5 & 112 & 62.5 & 212 \\
$\Psi$ [$\mu$rad\,s$^{-2}$] & 62.5 & 112 & 62.5 & 212 \\
\br
\end{tabular}
\end{indented}
\end{table}

As shown in Fig. \ref{fig_detect} the 1027Hz servo-loop provides the voltage to control the test-mass. This voltage is picked up at the output of the digital loop, filtered to prevent aliasing and downsampled to 4 Hz \cite{touboul12}. It is then multiplied by the a priori physical gain, delivered to the on-board computer (OBC) and sent to Earth for analysis.

\begin{table}
\caption{\label{tab_resbias} Laboratory measured resolution, bias, thermal sensitivity and gain of the capacitive sensors with a 100kHz applied voltage on the test-mass, $V_d=5\, Vrms$.}
\begin{indented}
\item[]\begin{tabular}{@{}lllllll}
\br
& $x$ & $y$ & $z$ & $\phi$ & $\theta$ & $\psi$ \\
\mr
Capacitive sensor's resolution \\
at $10^{-2}$Hz [$\mu$V\,Hz$^{-1/2}$] \\
IS1-SUREF & 8.1 & 3.8 & 3.7 & 9.3 & 3.8 & 3.7 \\
IS2-SUREF & 5.4 & 2.1 & 2.1 & 9.6 & 2.1 & 2.1 \\
IS1-SUEP & 12 & 3.1 & 3.1 & 12 & 3.1 & 3.1\\
IS2-SUEP & 5.6 & 1.9 & 1.9	 & 5.5 & 1.9 & 1.9\\
\mr
Bias [V] \\
IS1-SUREF & -0.006 & -0.007 & -0.001 & 0.014 & 0.002 & -0.003 \\
IS2-SUREF & 0.014 & -0.001 & -0.001 & -0.046 & 0.000 & 0.001\\
IS1-SUEP & 0.014 & -0.001 & -0.009 & 0.049 & 0.001 & -0.004\\
IS2-SUEP & 0.021 & -0.002 & -0.005 & 0.030 & 0.001 & -0.001 \\
\mr
Thermal sensitivity [$\mu$V/K] \\
IS1-SUREF & 64.7 & 10.1 & 14.3 & 36.6 & 1.7 & -2.1\\
IS2-SUREF & 16.7 & 12.0 & 25.4 & 130.9 & 2.7 & 12.2 \\
IS1-SUEP & 65.9 & 23.3 & 10.7 & 37.7 & 10.4 & -0.6 \\
IS2-SUEP & 16.3 & 8.9 & 14.0 & 193.3 & 1.2 & -0.5 \\
\mr
Capacitive sensor's gain [V/pF] \\
IS1-SUREF & 82.5 & 16.9 & 17.3 & 82.2 & 16.9 & 17.3\\
IS2-SUREF & 40.6 & 5.0 & 5.2 & 84.5 & 5.0 & 5.2\\
IS1-SUEP & 81.2 & 16.0 & 16.1 & 81.0 & 16.0 & 16.1 \\
IS2-SUEP & 39.3 & 5.0 & 5.0 & 85.0 & 5.0 & 5.0\\
\br
\end{tabular}
\end{indented}
\end{table}

Table \ref{tab_resbias} shows the measured resolution, bias, thermal sensitivity and gain of the capacitive sensing for all test-masses for all degrees of freedom. The capacitive sensor resolution and bias have been measured in the laboratory before flight, for each axis in open loop and are fully consistent with the objective of the mission: the WEP test at $10^{-15}$ accuracy.  
The capacitance gradient along the $X$-axis is fixed by the geometry, and the detector sensitivity along the $X$-axis is computed to 0.30 V/$\mu$m for all accelerometers; thence, the noise of the sensor corresponds to less than $4 \times{}10^{-11}$m\,Hz$^{-1/2}$. Along $Y$ and $Z$, the performance and sensitivity of the detector are of the same order. For the attitude motion of the test-mass, the sensitivity is estimated to be between $10^{-2}$V/$\mu$rad and $10^{-3}$V/$\mu$rad.
Because each capacitive sensing is used inside a servo-loop, its accuracy is not critical. The bandwidths of the sensors are sufficient and measured to $>160$\,Hz (-3dB), which are sufficient. With the measured thermal sensitivity of the electronics, a thermal stability of 1\,K\,Hz$^{-1/2}$ is required to achieve the resulting position measurement noise.

\begin{figure}
\begin{center}
\includegraphics[width=0.75\textwidth]{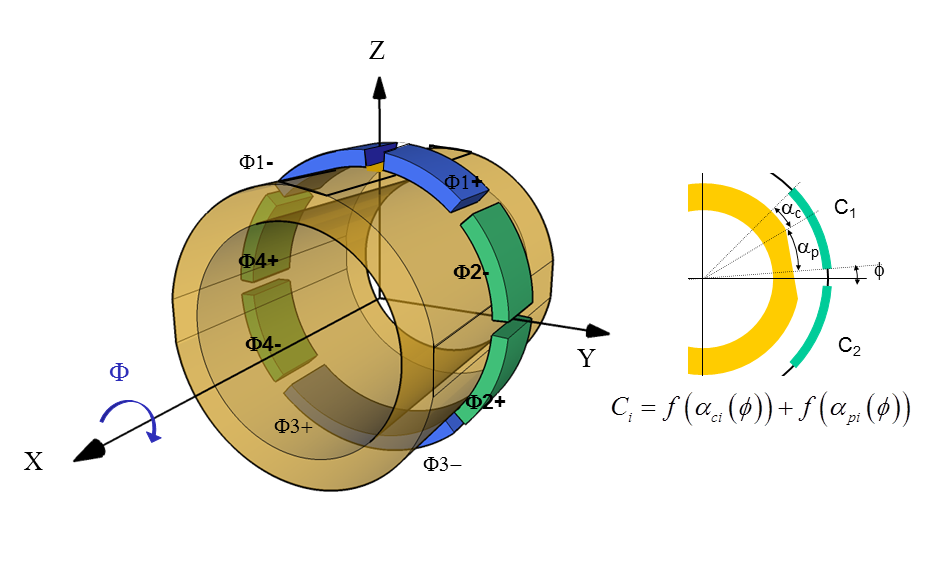}
\caption{Electrode arrangement for $\Phi$ measurement}
\label{fig_Phi}       
\end{center}
\end{figure}

\begin{figure}
\begin{center}
\includegraphics[width=0.45\textwidth]{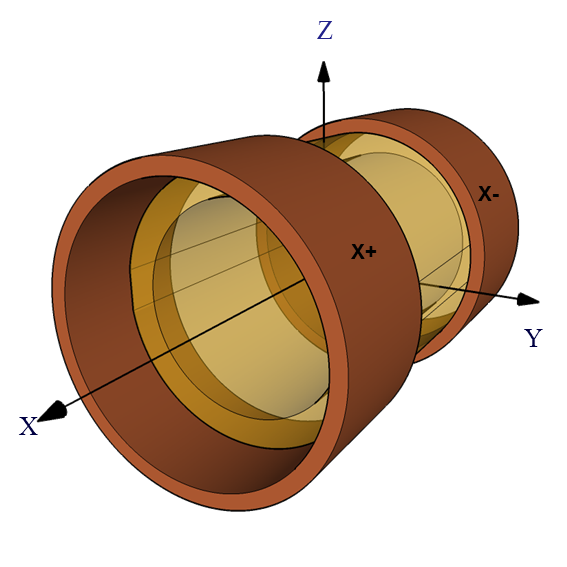}
\includegraphics[width=0.45\textwidth]{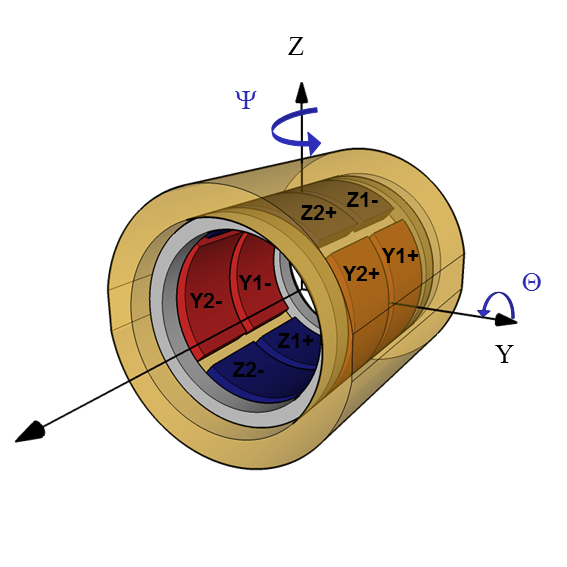}
\caption{Electrode arrangement for $X$ (left), $Y$, $Z$, $\Theta$ and $\Psi$ (right).}
\label{fig_xyz}       
\end{center}
\end{figure}

The electrode configuration used for test-mass control about its six degrees of freedom is shown in Figs. \ref{fig_Phi} and \ref{fig_xyz}.
The test-masses are controlled in translation and rotation. 
Independent pairs of electrodes are used to control the translation along $Y$ (and $Z$) using the mean value of the capacitive sensing given by the two pairs $Y1$ and $Y2$ (and $Z1$ and $Z2$). The rotation about $Z$ (and $Y$) uses the same set of electrodes $Y1$ and $Y2$ (and $Z1$ and $Z2$) but now the difference of the capacitive sensing is calculated.
For translation along $X$, the $X+$ and $X-$ pair of electrodes is used. Finally, for the rotation about $X$, a set of 8 electrodes have been electrically connected to form 2 assemblies of electrodes ($\Phi1-$ to $\Phi4-$ and $\Phi1+$ to $\Phi4+$) that are sensitive to the $\Phi$ motion. 
The actuation voltages on each electrode come from a drive voltage amplifier (DVA) and are calculated by the DSP that takes into account the 6 degrees of freedom capacitive sensing. The characteristics of each DVA have been verified on ground; in particular, we checked that the low-frequency DVA noise increases with a $f^{-1/2}$ law, below $3\times 10^{-2}$Hz.

For the $X+$ and $X-$ channels, each DVA has a matched gain of 16.000 to an accuracy $<0.2$\% for the particular case of the channel $X+$ and $X-$. The sensor outputs is from the PID controller. The satellite DFACS reduces the common mode acceleration, but the measurement output may be sensitive to actuator fluctuations and in particular the thermal ones (see Sect. \ref{sect_measure}). 

Additionally, each DVA command may be biased, giving two possible effects. If the same bias is applied on two electrodes which control  a test-mass degree of freedom, say $X+$ and $X-$, then it acts as a bias in the $V_p$ reference voltage (in the way as $V'_p$); alternatively, if it acts non symmetrically then it results in an offset in the applied restoring force along $X$.

Table \ref{tab_resbias2} lists the bias and thermal sensitivity of the actuator electronics (DVA) at electrodes $X+$ and $X-$ (that control translation along the $X$-axis) as measured in the laboratory before flight, for each test-mass.

The physical gain (that relates the electrode voltage to the measured electrostatic acceleration) along the $X$-axis of IS1-SUEP is estimated to be $6.89\times{}10^{-8}$ms$^{-2}$/V, while that of IS2-SUEP is estimated to be $8.05\times{}10^{-8}$ms$^{-2}$/V. In the same way, the physical gain along the $X$-axis of IS1-SUREF electrodes is estimated to be $6.89\times{}10^{-8}$ ms$^{-2}$/V and that of IS2-SUREF to $5.37\times{}10^{-8}$ms$^{-2}$/V.

\begin{table}[t]
\caption{\label{tab_resbias2} Bias and thermal sensitivity of the actuator electronics (DVA) at electrodes $X+$ and $X-$ (that control translation along the $X$-axis) as measured in the laboratory before flight.}
\begin{indented}
\item[]\begin{tabular}{@{}lll}
\br
& $X+$ & $X-$ \\
\mr
Noise (all masses) [$\mu$V\,Hz$^{-1/2}$] & 0.4 & 0.4 \\
\mr
Bias [$\mu$V] \\
IS1-SUREF &  157.05 & 167.56 \\
IS2-SUREF & 117.95 & 145.64 \\
IS1-SUEP & 146.80 & 224.86 \\
IS2-SUEP &  255.64 & 253.58 \\
\mr
Thermal sensitivity [$\mu$V/K] \\
IS1-SUREF & 0.32 & 1.00 \\
IS2-SUREF &  -0.78 & 0.86 \\
IS1-SUEP &  -2.01 & -1.11 \\
IS2-SUEP &  0.44 & -0.89 \\
\br
\end{tabular}
\end{indented}
\end{table}


\section{Launch and mission operations} \label{sect_commissioning}

MICROSCOPE was launched from Kourou on April 25, 2016 and injected into a sun-synchronous, circular, Low Earth orbit. Its mean semi-major axis 7090km and small eccentricity ($1.4 \times{}10^{-3}$) are perfectly compliant with science requirements. In particular, a low eccentricity reduces the disturbing effect of the Earth's gravity gradient at the $f_{\rm EP}$ frequency. The 710 km altitude was chosen from a trade-off to minimise the atmospheric drag while maximising the strength of the Earth's gravitational acceleration. A sun-synchronous orbit is beneficial to stabilise the temperature of the satellite. Given these orbital parameters, the orbital frequency $f_o = 1.6818 \times{}10^{-4}$ Hz.

\begin{figure}
\begin{center}
\includegraphics[width=0.75\textwidth]{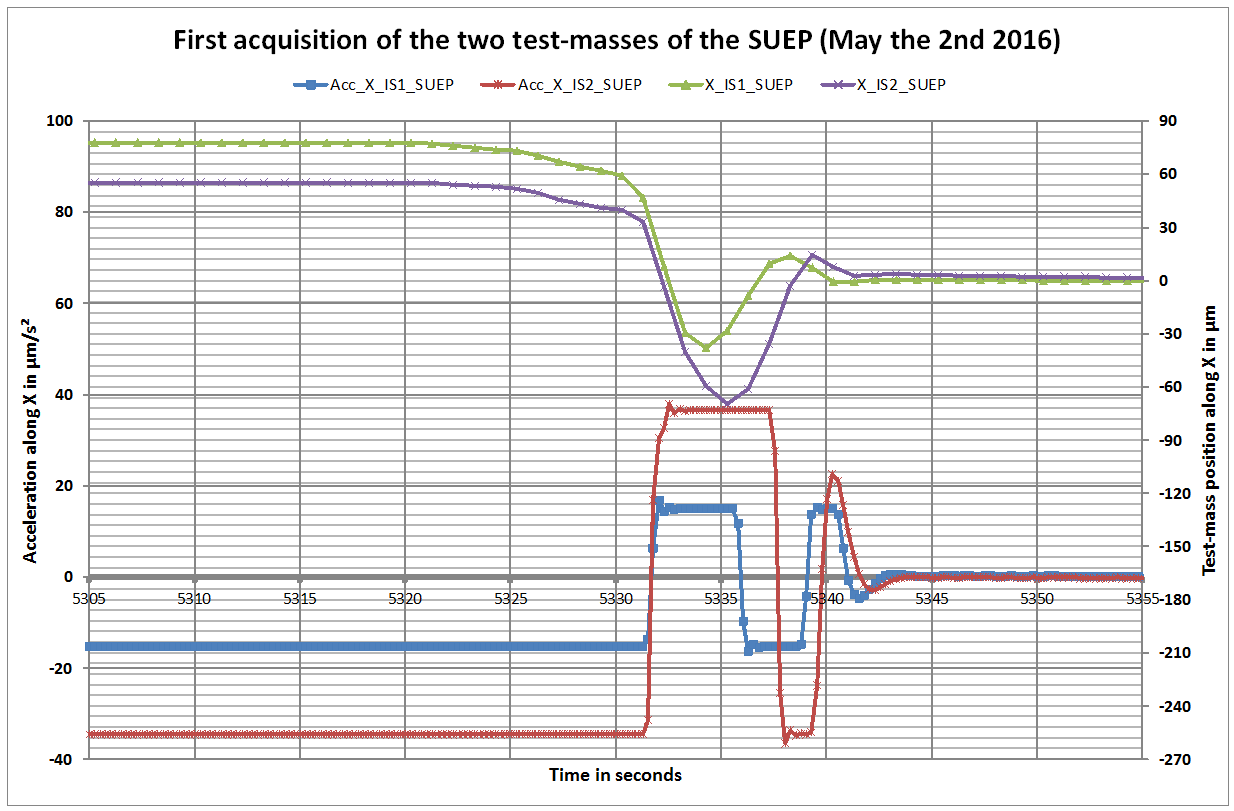}
\includegraphics[width=0.75\textwidth]{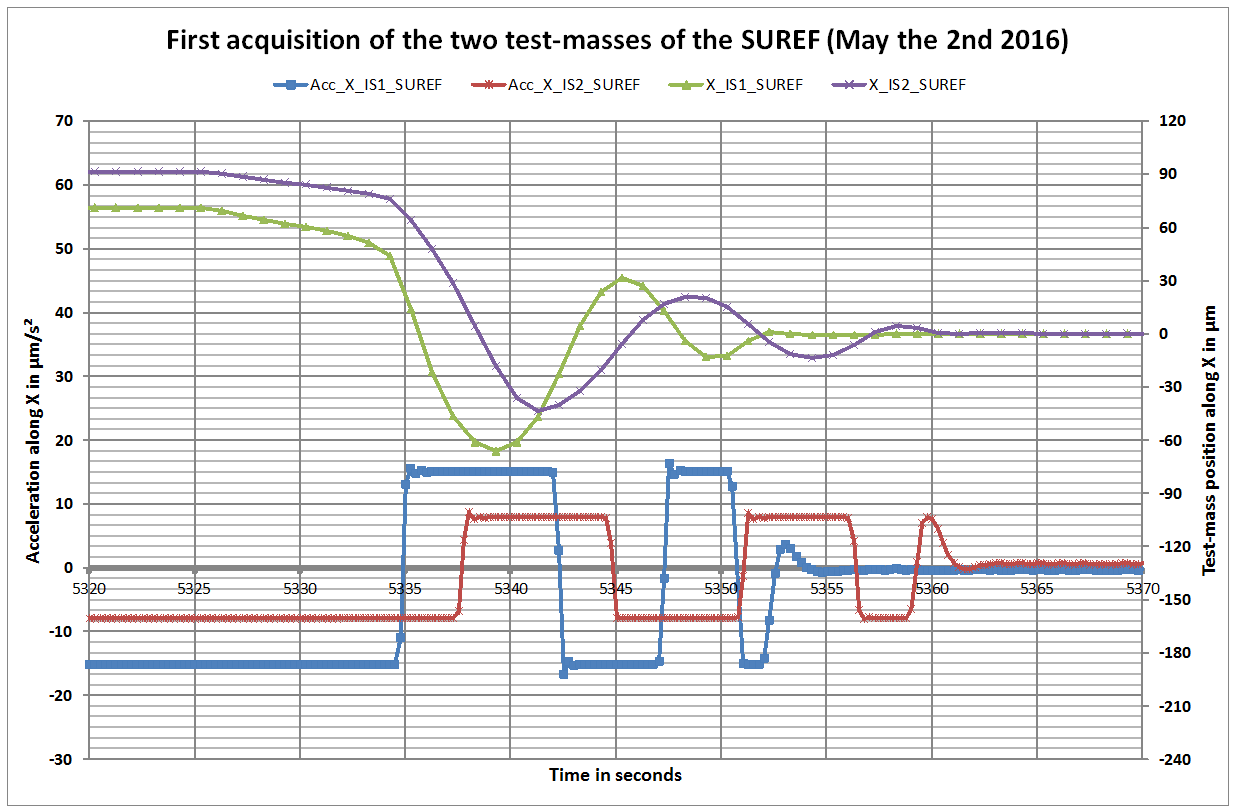}
\caption{Transient phase during the first test-mass levitation acquisition; before the release of the test-masses, all detectors show the test-masses locked on the stops; after release, the position sensor for each degree of freedom is controlled to null; the operation is perfectly autonomous from the first levitation on May, the 2nd. Both Sensor Units are shown. In both panels, green and purple lines represent the position of the test-mass along X, and the blue and red lines show their control electrostatic acceleration along X.}
\label{fig_TMrelease}       
\end{center}
\end{figure}

The T-SAGE instrument was switched on one week later (May 2nd, 2016) and its four test-masses were levitated about their six degrees of freedom. 
Fig. \ref{fig_TMrelease} shows the measured electrostatic acceleration and position of SUEP (upper panel) and SUREF (lower panel) of all four test-masses during release at the same time. Before release, the test masses are locked. Therefore the electrostatic control exerts its maximum force leading to a saturated measured acceleration. The position measurement gives the locked position. When the test-masses are unlocked, they start to oscillate and their position converges at the centre of the cage. They are then acquired and controlled by the electrostatic servo-loop.

After a brief verification of the satellite's behaviour, where all its operational modes were checked, the DFACS was turned on and the six degrees of freedom of the satellite were continuously servo-controlled with the help of the measurements provided by the scientific payload and the star-trackers. 
The DFACS provides a very soft acceleration environment to the experiment \cite{prieur17}. 

After the commissioning phase ended on November 14, 2016, the satellite and payload were declared ready for science operations in optimal thermal environment conditions. Since then, the science program has been managed as a succession of independent sessions in order to allow in-orbit flexibility of the mission scenario which can be modified weekly. Several science sessions have been successively performed with SUREF and SUEP: in-orbit calibration sessions of five orbits in inertial pointing and WEP test sessions of 120 orbits with the satellite spinning about its axis normal to the orbital plane. 
In this paper, as in Ref. \cite{touboul17}, we focus on only two WEP test sessions: one with the SUREF instrument of 62 orbits and one of with the  SUEP instrument of 120 orbits. 

The science phase includes several measurement sessions dedicated either to the SUEP or the SUREF instrument. Because of a failure in a capacitor, the power consumption increased in the SUREF, and thus so did the operating temperature also.  In order to minimise the risk of a failure propagation, SUEP and SUREF were not used simultaneaously. For each Sensor Unit, the test-mass motions are compared by calculating the difference of electrostatic acceleration during up to 120 orbits, sampled at 4Hz rate. This duration was defined prior to  launch and is actually limited by the operation of the attitude controller that must be reset periodically. This limits the effect of any stochastic disturbance (instrument noise, stochastic distribution of accelerometric environment). The scenario alternates EP sessions and calibration sessions in order to monitor the stability of the experiment.

\begin{figure}
\includegraphics[width=0.95\textwidth]{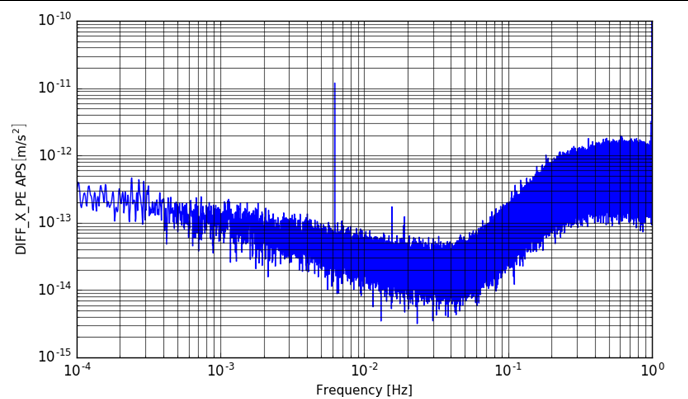}
\caption{Fast Fourier Transform (FFT) of the difference of the measured electrostatic acceleration along X during the scientific session with the SUEP instrument; $f_{\rm EP} = 3.1113 \times{}10^{-3}$ Hz; $f_o = 1.6818 \times{}10^{-4}$ Hz; satellite rotation frequency $f_s$ = $2.9432 \times{}10^{-3}$ Hz; the main peak caused by the gravity gradient is at  $2 f_o = 6.222 \times{}10^{-3}$ Hz. The peaks at higher frequencies are common mode signals that disappear after matching the scale factors.} 
\label{fig_ggt}       
\end{figure}

Moreover, we have defined the duration of the sessions as a multiple of orbital periods $T_o$, spin periods $T_s$ and EP periods $T_{\rm EP}$. This has the advantage of  getting a natural de-correlation between signals at multiples of $f_o$, $f_s$ and $f_{\rm EP}$. This is achieved by first estimating the orbital frequency from the orbit determination and then controlling the spin frequency to be a rational number times the orbital frequency. 
The spin frequencies have been selected to take advantage of the actual instrument levels and shapes (lower noise at higher frequencies: see Fig. \ref{fig_ggt}), as proposed also in the STEP mission \cite{sumner2007}: enforcing the compatibility between the spacecraft, the instrument capabilities and the natural de-correlation led to $f_s = 35/2 f_o$ for most sessions dedicated to the SUEP instrument, and $f_s = 9/2 f_o$ for most sessions dedicated to the SUREF instrument.

The rotation of the satellite is performed about the axis normal to the orbital plane, in the opposite direction to the orbital motion. Thus the apparent rotation of the Earth in the satellite frame defines the measurement frequency $f_{\rm EP}$ as the sum of the orbital frequency $f_o$ and of the satellite spin frequency $f_s$, $f_{\rm EP}=f_o+f_s$:
\begin{itemize}
\item $f_{\rm EP} = 3.1113 \times{}10^{-3}$ Hz  for SUEP in this paper;
\item $f_{\rm EP} = 0.92500 \times{}10^{-3}$ Hz for the SUREF in this paper.
\end{itemize}
In practice, the theoretical relation between the different periods mentioned above cannot be perfectly satisfied. Hardy et al. \cite{hardy13b} have studied realistic cases. In particular, with a specified error of $3\times{}10^{-8}\rm\,rad\,s^{-1}$ in the actual spin frequency, the projection rate of signals at frequencies $(n_1 f_o+n_2 f_s)$ over the $f_{\rm EP}$  frequencies does not exceed $10^{-4}$ (see Table 1 in). It has been checked in flight that this specification on the spin frequency is fully respected.

Tests of SUEP or SUREF at different frequencies have been performed since then and are being processed.

\section{Measurement, WEP signal and systematics} \label{sect_measure}

\begin{figure}
\center
\includegraphics[width=0.5\textwidth]{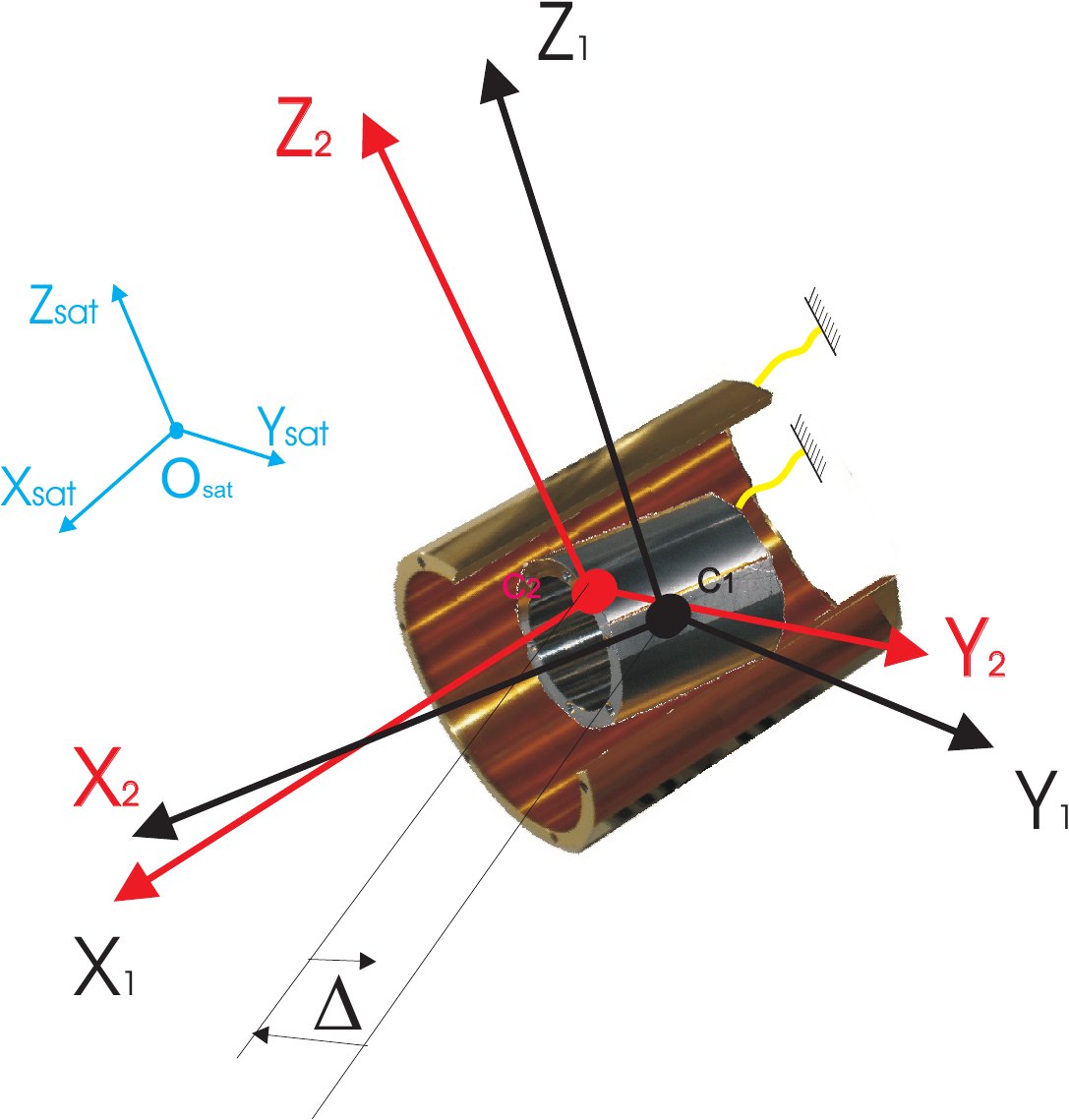}
\caption{Orientation of the test-mass axes versus satellite axes.}
\label{figor}       
\end{figure}

We define ${\vec \Gamma_k} $ as the acceleration exerted on the $k$-th test-mass by the electrodes that surround it. The three components of each acceleration  ${\vec \Gamma_k} $ are measured in the frame ($X_k$,~$Y_k$,~$Z_k$) attached to the corresponding test-mass electrode set (see Fig. \ref{figor}). Because of small (time-independent) misalignments with respect to the satellite frame,  ($X_{\rm sat}$,~$Y_{\rm sat}$,~$Z_{\rm sat}$), the locally measured components ${\vec \Gamma_k} $ are related to their components ${\vec \Gamma_k^{\rm sat}} $ in the satellite frame via
${\vec \Gamma_k}  = [\theta_k] {\vec \Gamma_k^{\rm sat}} $,  where the matrix $[\theta_k]$ reads 
\begin {equation}
\left[\theta_{k}\right] = \left[
\begin{matrix}
1 & \theta_{kz} & -\theta_{ky} \\
-\theta_{kz} & 1 & \theta_{kx} \\
\theta_{ky} & -\theta_{kx} & 1 \\
\end{matrix}
\right].  
\end {equation}
The three (antisymmetric) off-diagonal elements $\theta_{kl}$ measure the small
 rotation  between the satellite frame and the $k$-th test-mass frame ($\theta_{kl} < 2.5 \times{}10^{-3}$\,rad, as constrained by the construction of the MICROSCOPE instrument and its installation on board the satellite).
  
In addition to the antisymmetric off-diagonal elements $\theta_{kl}$, there are also other defects to be taken into account in the measurement equation: the control acceleration offsets, the non-unit scale factors (1+$K_{kl}$) and the couplings ($\eta_{kl}$). The measurement is then written as ${\vec \Gamma_k}^{\rm meas} = [A_k] {\vec \Gamma_k}$ where the sensitivity  matrix $[A_k]$ reads 
\begin {equation}
\left[A_k \right]~=~\underbrace{\left[
\begin{matrix}
1+K_{kx} & 0 & 0 \\
0 & 1+K_{ky} & 0 \\
0 & 0 & 1+K_{kz}
\end{matrix}
\right]}_{\text{scale factor}} + \underbrace{\left[
\begin{matrix}
0 & \eta_{kz} & \eta_{ky} \\
\eta_{kz} & 0 & \eta_{kx} \\
\eta_{ky} & \eta_{kx} & 0
\end{matrix}
\right]}_{\text{coupling}}.
\end {equation}

Any WEP violation will appear in the difference of accelerations between the inner mass ($k=1$) and the outer mass ($k=2$) of the SUEP sensor measurement, say
${\vec \Gamma_d}^{\rm meas} \equiv {\vec \Gamma_1}^{\rm meas} - {\vec \Gamma_2}^{\rm meas}$ (we call acceleration the ratio between the electrostatic force and the inertial mass). The derivation of the measurement is detailed in Ref. \cite{hardy13a}:

\begin{multline} \label{eq_gammad}
\overrightarrow\Gamma_d^{meas}\simeq \left[M_c\right]\left( \delta\left(2,1\right) \overrightarrow g \left( O_{sat} \right) +  \left( \left[ T \right]- \left[\rm{In}\right]  \right)\overrightarrow\Delta -
2\left[\Omega \right]\dot{\overrightarrow\Delta}-\ddot{\overrightarrow\Delta} \right) + \overrightarrow K_{0,d} \\
+ 2\left[ M_d\right] \overrightarrow\Gamma_c^{app} + 
\overrightarrow \Gamma_d^{quad} + \left[ \rm{Coupl}_d\right] \dot{\overrightarrow \Omega}+ \overrightarrow \Gamma_d^{n},
\end{multline}
where all quantities are expressed in the instrument frame and $\delta(2,1)$ is the potential WEP violation signal (approximate E\"otv\"os parameter of the outer mass (2) with respect to the inner mass (1) --see Sect. \ref{ssect_delta}) coupled to the Earth gravity acceleration vector in the satellite frame $\overrightarrow g(O_{\rm sat}) =\left(g_x, g_y, g_z \right)^T$. Other terms on the right-hand-side of the equation are instrumental and nuisance contributions to the measurement, which impact its accuracy and precision. They can be sorted into three main non-exclusive categories: (i) geometrical and mechanical imperfections, (ii) perturbative accelerations and (iii) electronic noise. We briefly list them here, before giving more details about their effects below.

Geometrical imperfections come from tiny differences in the centring, alignment and parallelism of the test-masses or the electrodes with respect to each other and to the satellite. The most obvious is the test-masses off-centring (their centres of mass are not exactly coincident): $\overrightarrow \Delta = \left(\Delta x,\Delta y,\Delta z \right)^T$ is the vector (in the satellite frame) connecting the centre of the inner mass to that of the outer mass (see Sect. \ref{ssect_ggt}). Their first and second time derivatives $\dot{\overrightarrow \Delta}$ and $\ddot{\overrightarrow \Delta}$ are nullified in the instrument's bandwidth when the instrument servo-controls maintain the masses motionless versus the satellite frame.
The off-centrings are coupled to the Earth gravity gradient tensor and to the matrix gradient of inertia (expressed in the satellite frame) $\left[ T \right]$ and $[\mathrm{In}]= \left[\dot \Omega\right] + \left[ \Omega \right]\left[ \Omega \right]$, creating a characteristic signal at the $2f_{\rm EP}$ frequency (see Sect. \ref{ssect_ggt}). The first derivative of the off-centring couples to the satellite angular velocity to give rise to a Coriolis effect $2\left[ \Omega \right]\dot{\overrightarrow \Delta}$ ; it is very weak because the relative velocity of the test-masses at the test frequency is limited by the integral term of the accelerometer's servo-loops and because the angular velocity is well controlled by the satellite DFACS loops.
Additionally, correlations in the accelerations projected on different axes, as well as projections of undesired contributions may result from misalignments, thereby contaminating the measurement. Those imperfections are accounted for in the common-mode and differential-mode sensitivity matrices $\left[M_c\right] = \frac{1}{2}\left(\left[A_1\right]\left[\theta_1\right]+\left[A_2\right]\left[\theta_2\right]\right)$ and
$\left[M_d\right] = \frac{1}{2}\left(\left[A_1\right]\left[\theta_1\right]-\left[A_2\right]\left[\theta_2\right]\right)$.  The $\left[M_d\right]$ matrix can be calibrated in flight to minimise their effect on the measurement (see Sect. \ref{ssect_calib}).

Mechanical imperfections impact the control of the test-mass positions. The mechanical parts must be well designed and integrated (with no residual free-motions, with stabilities of the instrument assembly, with low residual stiffness between the masses and these assemblies). The accuracy of the measurement is limited by the sensors position noise, by the inertial sensors servo-channel qualities and the stabilities of the instrument mechanics which form the instrument frame reference before launch, and by the performance of the servo-channel electronics that has been measured in the laboratory as already shown in Tables \ref{tab_resbias} and \ref{tab_resbias2}.

The most obvious disturbing accelerations are non-gravitational accelerations applied to both test-masses through the satellite (such as atmospheric drag and solar radiation pressure). They can be minimised by the satellite's drag-free system (see Sect. \ref{ssect_dragfree}).
As they are felt by both test masses, those accelerations are combined in the common-mode acceleration $\overrightarrow\Gamma_c^{app}$.
Additionally, $\overrightarrow \Gamma_d^{quad}$ is the difference of the non-linear terms in the measurement (mainly the difference of the quadratic responses of the inertial sensors). Other disturbing accelerations, that do not appear directly in Eq. (\ref{eq_gammad}) are due to: 

\begin{itemize}
\item radiation pressure, radiometer effect and residual gas damping --Sect. \ref{ssect_therm};
\item local gravity of the satellite --Sect. \ref{ssect_field};
\item magnetic field effect --Sect. \ref{ssect_field};
\item electric field --Sect. \ref{ssect_field};
\end{itemize}

Finally, Eq. (\ref{eq_gammad}) takes into account the vector $\overrightarrow K_{0,d}$ of the difference of the inertial sensor measurement offset, the matrix $\left[ \mathrm{Coupl}_d \right]$ of the difference, between the two sensors, of the coupling from the angular acceleration $\dot{\overrightarrow \Omega}$ to the linear acceleration (see Sect. \ref{ssect_coupling}), and the difference $\overrightarrow \Gamma_d^{n}$ of the acceleration measurement noises of the two sensors (coming from thermal noise, electronic noise, parasitic forces,...), including stochastic and systematic error sources.

\subsection{WEP signal} \label{ssect_delta}

$\delta \overrightarrow{g}$ (7.9\,m\,s$^{-2}$) is the signal to be possibly found if the WEP is violated to a high enough level. 
In an ideal experiment, the SUREF should give a null value while the SUEP gives a signal proportional to the E\"otv\"os parameter for that combination of materials. 

Since the $X$-axis is much more sensitive than $Y$ and $Z$, we use only measurements along this axis to look for an EP violation; the corresponding model is Eq. (\ref{eq_gammad}) projected on the $X$-axis. Thence, only the first lines of the matrices $[M_c]$, and $[M_d]$ are relevant to our analyses.

\subsection{Effects of off-centrings and gravity gradients} \label{ssect_ggt}

As shown by Eq. (\ref{eq_gammad}), the differential measurement is sensitive to the Earth gravity gradient, mainly modulated in the instrument frame at $2f_{\rm EP}=2(f_o+f_s)$  \cite {touboul12}. The amplitude of this signal depends on the off-centring $\overrightarrow \Delta$ between the centre of the test-masses.  We can easily see the corresponding peak in the frequency domain (at $2f_{\rm EP} = 6.222 \times{}10^{-3}$ Hz) in Fig. \ref{fig_ggt}.

We follow Ref. \cite{metris98} to compute the Earth gravity gradient tensor projected into the instrument frame, with the help of the measured position and attitude of the satellite and the ITSG-Grace2014s gravity potential model \cite{mayer06} expanded up to spherical harmonic degree and order 50. The distance between the two test masses' centres of mass is a priori unknown but its components along the $X$- and $Z$-axes (in the instrument frame, fixed to the satellite  frame) can be precisely estimated from the gravity gradient signal at $2 f_{\rm EP}$. \\
The contribution of this effect in the differential measurement can then be corrected; the remaining error after correction can be expressed by:
\begin{equation} \label{eq_dggt}
\left( [T] \overrightarrow{\Delta}_{DC} - \hat{[T]}\hat{\overrightarrow{\Delta}}_{DC} \right)  \approx  \left( [T]- \hat{[T]} \right) \overrightarrow{\Delta}_{DC} + \hat{[T]} \left( \overrightarrow{\Delta}_{DC} - \hat{\overrightarrow{\Delta}}_{DC} \right) 
\end{equation}
where $\hat{\alpha}$ denotes the estimate of $\alpha$.
The first term is due to  error  on the gravity gradient, expressed in the instrument frame, used to correct the effects of the estimated off-centring. The second term is due to the error of calibration of the off-centring. In contrast with the STEP mission \cite{sumner2007}, it is better here not to correct the real position of the test-mass to cancel the gravity-gradient effects. Indeed, this position corresponds to the zero of the capacitive sensor that is not optimised to operate far from this position. In closed loop, the capacitive sensor output is null because an equivalent force is applied to displace the test-mass in order to nullify the output of the capacitive sensing. This back-action force turns into an offset in the accelerometer measurement output. 

Given the very small distance between the two test-masses, an error on the gravity gradient limited to $10^{-11}$s$^{-2}$ leads to an error smaller than a few $10^{-16}$ms$^{-2}$ on the acceleration correction. The intrinsic knowledge of the Earth gravity potential ensures an error  much smaller than  $10^{-11}$s$^{-2}$ on the gravity gradient tensor; however it is also necessary to know the position of the satellite and its attitude (to convert the gradient tensor from the Earth frame to the instrument frame) with a sufficient precision.
The precise orbit and attitude are provided by CNES. The orbit determination is based on a Doppler tracking system currently used in the Myriad Satellite Line and  on-board Global Positioning System (GPS) receiver measurements. For the SUEP session, the accuracy of the position knowledge is estimated to be 0.061 m along the radial direction, 0.109 m along the cross-track direction and 0.133 m along the tangential direction; this is much better than needed (the most stringent requirement is specified to 7 m).

The attitude of the satellite is evaluated by filtering and combining  the on-board star tracker system outputs and the angular acceleration measurements provided by the instrument itself. The alignment between the star tracker frame (satellite frame) and the instrument frame has been calibrated by ground measurements during the satellite integration and after qualification. This ground calibration is used to project the star sensor frame onto the instrument one but the small misalignment values ($\approx 10^{-4}$ rad), allow it to be neglected. The hybridisation of the accelerometer and the star sensor measurements create a systematic error depending on frequency. When the satellite rotates, the star sensor exhibits an accuracy of 0.14 $\mu$rad about $X$-axis (instrument frame), 0.81 $\mu$rad about $Y$-axis and 0.13 $\mu$rad about $Z$-axis. These accuracies are compliant with the correction of the gravity gradient to $10^{-16}$ ms$^{-2}$ (requiring only 1$\mu$rad).

To mitigate uncertainties in the off-centring, we benefit from the fact that the gravity gradient signature is mainly at $2f_{\rm EP}$ frequency, allowing $\Delta x$ and $\Delta z$ to be estimated from this signal; its contribution at $f_{EP}$ is small in inertial mode and completely negligible in spinning mode (which is the case considered here) \cite{touboul12}.
In the nominal configuration, where the $Y$-axis is normal to the orbital plane, the gravity gradient due to the off-centring along $Y$ is negligible. Nevertheless, it is calibrated (and corrected if necessary) through a dedicated session, where we project the gravity gradient along the $X$-axis by biasing the satellite star tracker output which causes the DFAC to swing the satellite.
The off-centring has been estimated for the SUEP during the EP session (120 orbits) for the $X$ and $Z$ components and during the calibration session after the EP session (5 orbits) for the $Y$ component:
\begin{itemize}
\item Along $X$: $\Delta x = 20.14 \pm{}0.05 \mu$m
\item Along $Z$: $\Delta z = -5.55 \pm{}0.05 \mu$m
\item Along $Y$: $\Delta y = -7.4 \pm{}0.2 \mu$m.
\end{itemize}

\subsection{Calibration} \label{ssect_calib}

Dedicated sessions are used for in-flight calibration: stimuli specific to each parameter are applied in the DFACS satellite loop \cite{guiu07,jafry02} or in the test-mass control loop \cite{hardy13a, hardy12}. The calibration allows us to match both the sensitivities of the sensor and the alignments of their $X$ axes and to verify the quadratic term levels \cite{hardy13a, guiu07, hardy12}. Note that it is designed to optimise the precision of the measured acceleration along the most sensitive axis ($X$).

\begin{figure}
\includegraphics[width=0.95\textwidth]{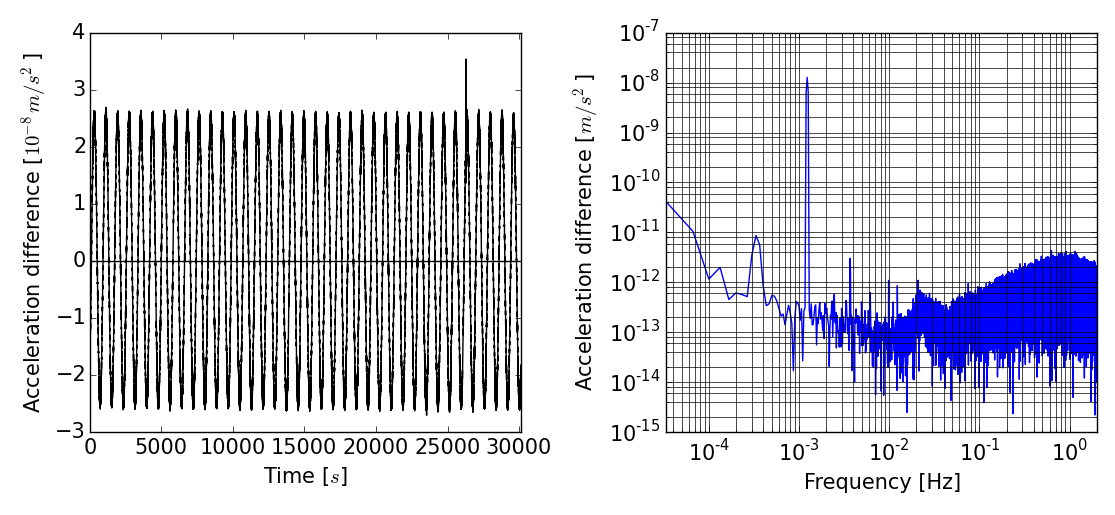}
\caption{Difference of acceleration measured along the $X$-axis during a $[M_d]$ matrix calibration session, in the time domain (left) and frequency domain (right). A sine wave of $5 \times{}10^{-8}$ms$^{-2}$ at $1.2285 \times{}10^{-3}$Hz was added to the drag-free loop. The spike about $t=26000$ s is due to a transient acceleration applied to the satellite and seen through the transfer function and the anti-aliasing filter of the measurement channels; a residual signal remains in the difference; such a large signal occurs less than once a week and is probably due to micro-debris.}
\label{fig_calib}       
\end{figure}

In order to calibrate some elements of the matrix ${M_d}$, a sine wave linear acceleration of $5 \times{}10^{-8}$ms$^{-2}$ at the frequency $f_{\rm calib} = 1.2285 \times{}10^{-3}$Hz is applied to the satellite propulsion by biasing the SU measurement output used by the DFACS along one axis (Fig. \ref{fig_calib}). For a given SU, both test-masses undergo the same acceleration, allowing for the estimate of the difference of their sensitivities along their various axes, their misalignment and their cross-axes coupling (having previously demonstrated a sufficient sensor output linearity). 

\begin{table}[t]
\caption{\label{tab_calib} Coefficients of the first line of the $[M_d]$ matrix, as estimated in orbit with dedicated sessions. The quadratic terms $K_{2i}$ were evaluated by exploiting the $2\times{}f_{\rm calib}$ frequency during the same sessions.}
\begin{indented}
\item[]\begin{tabular}{@{}lll}
\br
Parameter & SUREF & SUEP \\
\mr
$M_{d11}/M_{c11}$ & $-1.43\times{}10^{-2} \pm{}0.8\times{}10^{-4}$ & $8.56\times{}10^{-3}\pm{}6.5\times{}10^{-5}$ \\
$M_{d12}/M_{c11}$ [rad] & $-2.62\times{}10^{-5} \pm{}2.4\times{}10^{-6}$ & $-2.63\times{}10^{-4} \pm{}6.4\times{}10^{-6}$ \\
$M_{d13}/M_{c11}$ [rad] & $-8.90\times{}10^{-5} \pm{}2.0\times{}10^{-6}$ & $1.24\times{}10^{-4} \pm{}1.1\times{}10^{-5}$  \\
$(K_{21} - K_{22})/ M^2_{c11}$ [m$^{-1}$s$^2$] & $1795 \pm{}82$ & $695 \pm{}335$  \\
$|K_{21}|$ [m$^{-1}$s$^2$] & $\leqslant 3800$ & $\leqslant 900$ \\
$|K_{22}|$ [m$^{-1}$s$^2$] & $\leqslant 1500$ & $\leqslant 600$ \\
\br
\end{tabular}
\end{indented}
\end{table}

The requirements on both SU's $M_{d11}/M_{c11}$, $M_{d12}/M_{c11}$ and $M_{d13}/M_{c11}$ terms were established before launch, based on an analytical error budget. As shown in Table \ref{tab_calib}, they are virtually obtained by construction and integration of the instrument, without correction. Nevertheless, those terms are estimated in orbit just before each respective EP session, with accuracies better than $10^{-4}$ (two orders of magnitude better than the requirements).

Two methods were used to check the sensor linearity along their $X$-axis (i.e., estimate their quadratic terms). The first one is a by-product of $M_{d11}/M_{c11}$ calibration sessions: $(K_{21}-K_{22})/M^2_{c11}$ is readily extracted from the $2f_{\rm calib}$ signal.
The second method is based on applying a square wave acceleration to the satellite: a large amplitude 60Hz-signal acceleration ($1.3\times{}10^{-7}$m\,s$^{-2}$ to $2\times{}10^{-7}$ m\,s$^{-2}$ depending on the mass) is alternatively turned on and off during 500s phases. The servo-loop's gain rejects the response at 60Hz and 120Hz, while the quadratic response produces a constant signal added as an offset to the control acceleration during 500s every 1000s: the expected signal is thus a mHz-square signal proportional to the quadratic coefficient.
Due to the low values of the quadratic terms realised in-flight, the response to the stimuli is at the limit of the sensitivity, so that Table \ref{tab_calib} reports only upper bounds.
Both methods provide values much lower than requirements ($|K_{21}| <$ 20 000 m$^{-1}$s$^2$ and $|K_{22}| <$ 6 000 m$^{-1}$s$^2$) (Table \ref{tab_calib}).

\subsection{DFACS operation and impact} \label{ssect_dragfree}

Since the inertial sensor sensitivities are not perfectly identical (the sensitivity $[M_d]$ is not null), the difference of acceleration measurement is sensitive to the level of the platform's residual acceleration (Eq. \ref{eq_gammad}).

\begin{figure}
\center
\includegraphics[width=0.7\textwidth]{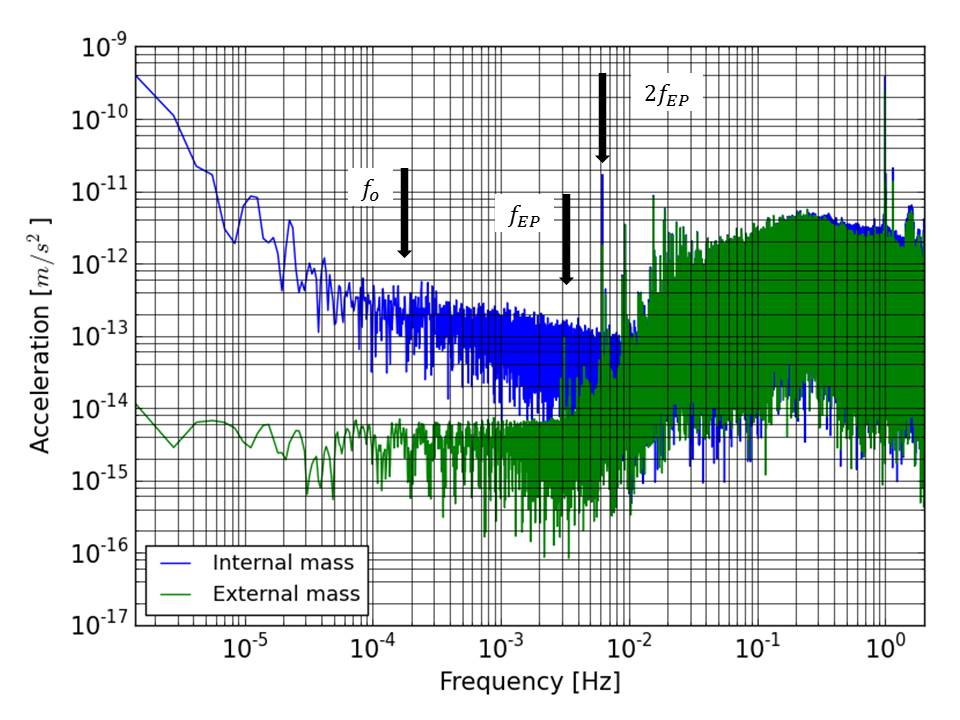}
\caption{FFT of the SUEP acceleration measurement along X for the inner mass (IS1 --blue) and the outer mass (IS2 --green) over 120 orbits. The IS2 test-mass output provides the measurements to the drag-free loop (which has a bandwidth of about 0.01 Hz).}
\label{fig_dragfree}       
\end{figure}

When in operation, the satellite's drag-free control acts on the propulsion system to cancel the measurement output of one test-mass (or more rarely of a combination of two test-masses). Then, the residual acceleration measured by the sensor controlling the drag-free compensation represents the residual of the DFACS pilotage.
Fig. \ref{fig_dragfree} shows the measured acceleration when the drag-free is controlled by SUEP's external mass: the DFACS residual is less than $10^{-14}$\,m\,s$^{-2}$ in the bandwidth of interest [$10^{-4}$ Hz -- $4\times10^{-3}$ Hz] (green line). The other sensor gives an upper bound of the residual acceleration experienced by the satellite. It mainly contains the residual common mode acceleration ($\leqslant 10^{-13}$\,m\,s$^{-2}$ at $f_{EP}=3.1\times10^{-3}$ Hz) and all systematic errors (the gravity gradient dominates at $2f_{EP}$). It is one order of magnitude smaller than the requirements, which helps to reduce the constraint on the calibration accuracy.

The DFACS also controls the satellite's attitude in order to limit the variation of its angular acceleration and velocity at $f_{EP}$. The resulting residuals are lower than $7 \times10^{-12}$ rad\,s$^{-2}$ and $3.6 \times10^{-10}$ rad\,s$^{-1}$, respectively.

\subsection{Instrument error analysis} \label{ssect_coupling}

\subsubsection{Noise characteristics}

\begin{figure}
\centering
\includegraphics[width=0.9\textwidth]{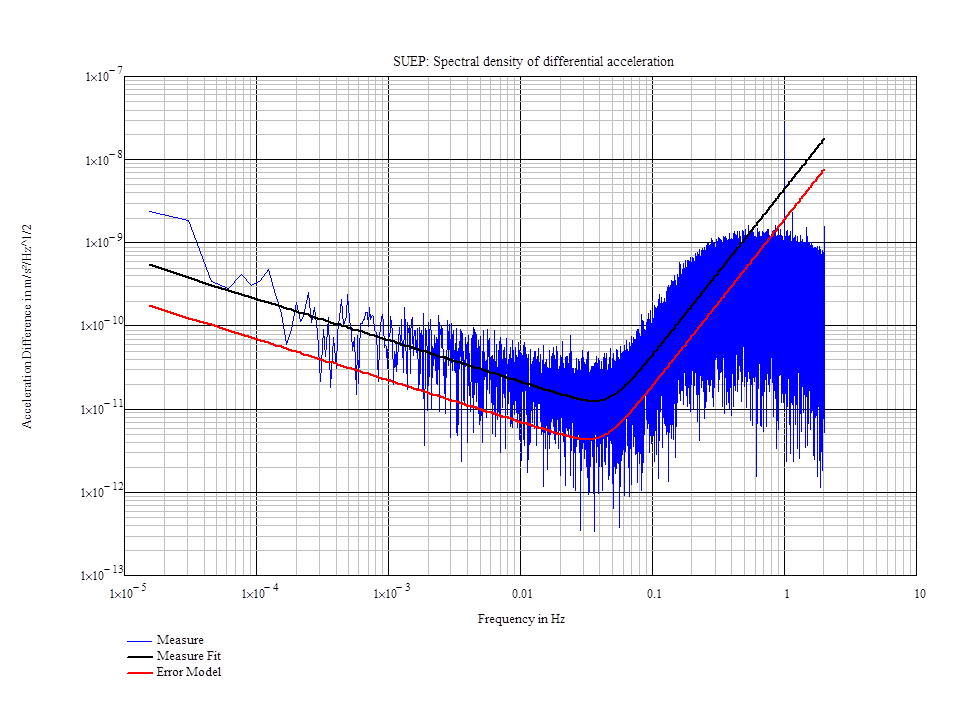}
\includegraphics[width=0.9\textwidth]{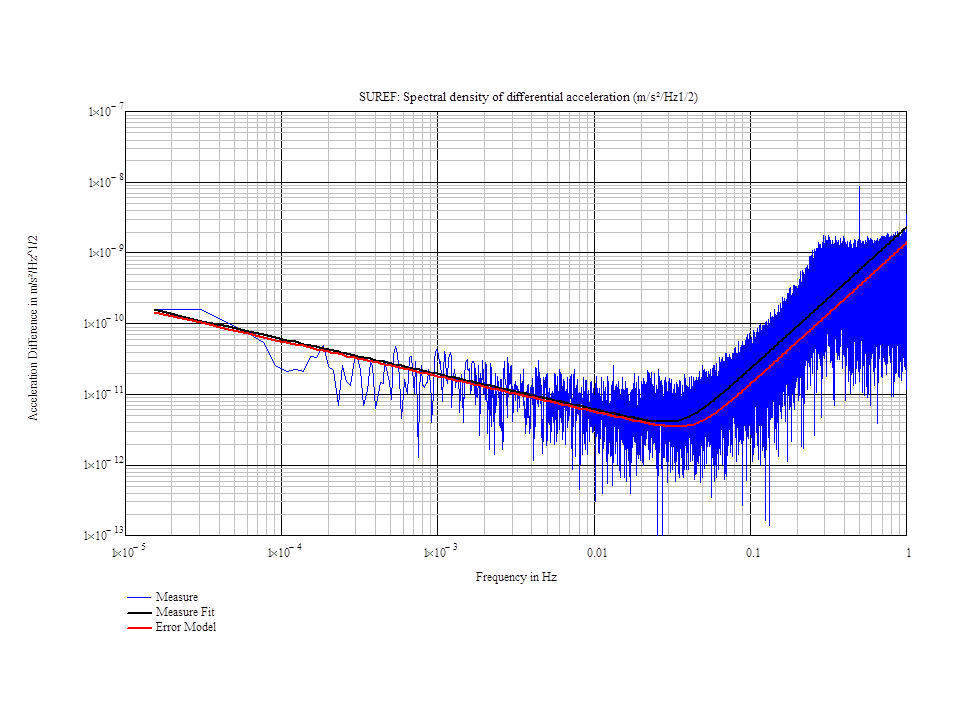}
\caption{Spectral Density of the sensor difference of acceleration along the x sensitive axis for SUEP (upper panel) and SUREF (lower panel): measured noise (in blue) with its fit (in black) and modelled before flight (in red).}
\label{fig_noise}       
\end{figure}

The frequency characteristics of the instrument noise were finely analysed before launch, both through tests on the MICROSCOPE flight models and through our experience from previous missions (GRACE, GOCE -- Refs.  \cite{touboul99, touboul04, flury08, marque10}). Each inertial sensor's error budget was established in Ref. \cite{touboul12}. Fig. \ref{fig_noise} shows the expected difference of acceleration noise along the $X$ axis (red curves) and compares it with the measured noise (blue curves). The deviation of the measured noise with respect to the expected one is not yet totally explained. The noise depends mainly on three terms, which we describe below. 

\begin{figure}
\center
\includegraphics[width=0.45\textwidth]{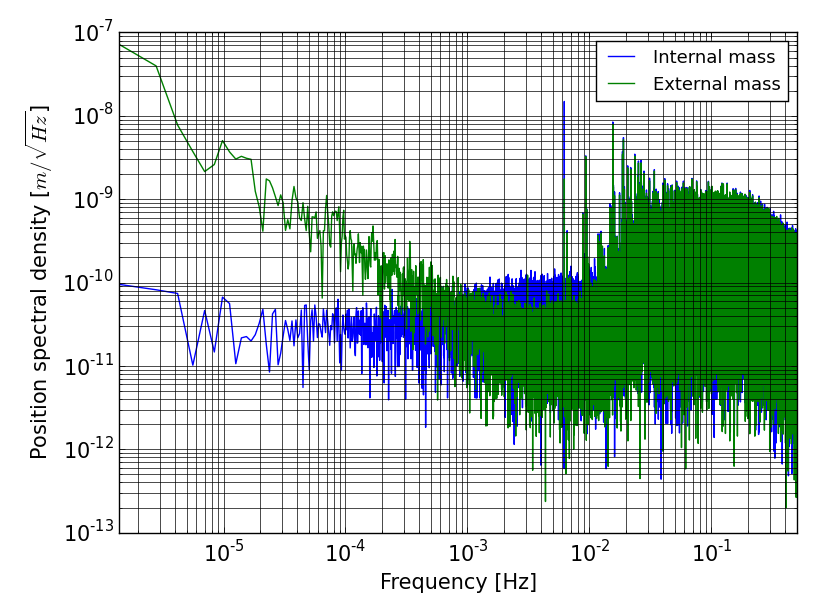}
\includegraphics[width=0.45\textwidth]{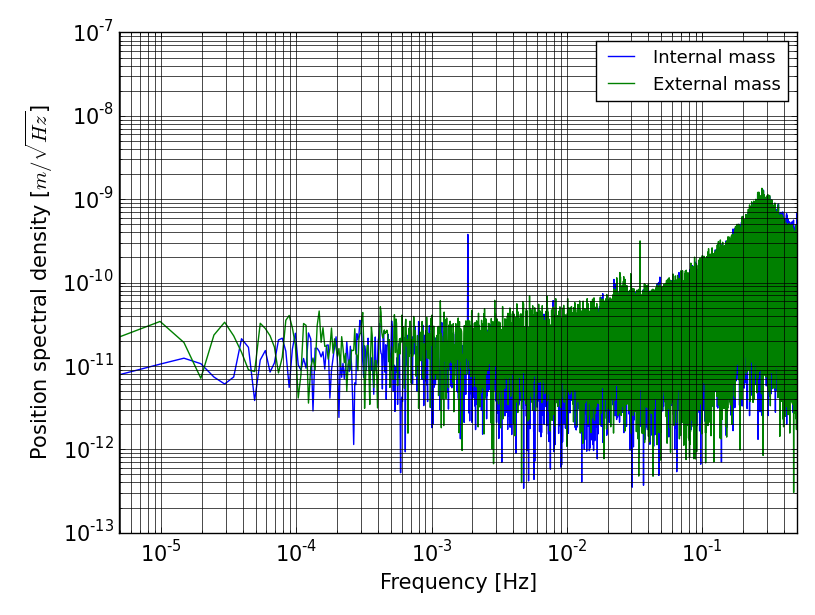}
\caption{Spectral density of the test-mass position measurement along X: SUEP (left) and SUREF (right)}
\label{fig_pos218}       
\end{figure}

At higher frequencies, the noise is dominated by the contribution of the position sensor noise multiplied by the square of the angular frequency. The position of the test-masses inside their electrode cage is measured by capacitive sensing converted into displacement, sampled at 1 Hz and stored in the housekeeping data. Fig. \ref{fig_pos218} shows the spectral density of the test-mass position measured along the $X$-axis for SUEP and SUREF. The maximum level of noise around $f_{\rm EP}$ and $2 f_{\rm EP}$ is less than $10^{-10}$\,m\,Hz$^{-1/2}$ for all four masses. 
The shape at low frequency of the inner test-mass position spectrum is determined by the PID control law and the level of acceleration. In particular in the experiment leading to Fig. \ref{fig_pos218}, we implemented a softer PID for SUEP than for SUREF. This new PID gave better rejection of aliasing of resonant frequencies between 10Hz and 12Hz into the range of 0.1Hz to 1Hz.
A large number of peaks from $6\times{}10^{-3}$Hz to $4\times{}10^{-2}$Hz in common-mode measurements almost completely disappear in the differential measurement.

Finally, the spectrum is governed by damping of the test-mass motion with respect to the instrument frame ($f^{-1/2}$ behaviour) at low frequencies, and by the thermal environmental noise ($f^{-1}$ behaviour) at very low frequency. The latter depends on the thermal sensitivity of the instrument and the environment temperature fluctuations.  The damping has been attributed to the gold wire used to control the charge of the test mass. This effect is supposed to be much greater than the one caused by residual gas in the vacuum vessel due to  out-gassing of mechanical parts : the use of silica parts and getter material limit the amount of residual gas.
Using a rough assumption about the quality factor ($Q=100$) of the gold wire \cite{willemenot00b}, Fig. \ref{fig_noise} shows the best log-log fits to the measured spectral density, with  known frequency laws $f^{-1/2}$ and $f^2$. The red curves correspond to the expected noise with the error model established before the launch.  Though it shows a good order of magnitude, the rough model may be improved. 

Table \ref{tab_bias} shows the values of the acceleration offset and noise (around $10^{-3}$ Hz) observed on the $X$-axis for all inertial sensors. 
The biases are computed as the mean values of the outputs over an integer number of orbits, without no drag-free. This is a good approximation since the drag acceleration averages to zero over one orbit, and the effects of the gravity gradients (Earth and satellite) and of the radiation pressure are negligible compared to the biases.
The values of the noise at $10^{-3}$\,Hz correspond to the maximum of the spectrum at this frequency. 

The noise levels on the $Y$ and $Z$ axes are observed to be less than a few $10^{-11}$\,m\,s$^{-2}$Hz$^{-1/2}$ and $10^{-10}$\,m\,s$^{-2}$Hz$^{-1/2}$ respectively. Given the  low cross-coupling ($<10^{-4}$) observed during calibrations, the noise sources from the $Y$ and $Z$ projected on the $X$-axis are negligible.

\begin{table}[t]
\caption{\label{tab_bias} Control acceleration offset and noise along $X$ of the four inertial sensors: a good approximation of the offsets can be given by the mean values of the acceleration measurement outputs over an integer numbers of orbits (when the drag-free is not operating), the noise is evaluated by the maximum values of the spectra of the sensor output difference at the stated  frequency.}

\begin{indented}
\item[]\begin{tabular}{@{}lllll}
\br
& Measured acceleration offset [m\,s$^{-2}$]  \\
\mr
SUREF internal mass (IS1)	& $-1.4\times{}10^{-7}$ \\
SUREF external mass (IS2)	& $ 7.7\times{}10^{-7}$	\\
SUEP internal mass (IS1) & $3.4 \times{}10^{-8}$  \\
SUEP external mass (IS2) & $-1.4\times{}10^{-6}$  \\
\br
& Observed acceleration noise at $10^{-3}$Hz [m\,s$^{-2}$Hz$^{-1/2}$] \\
\mr
SUREF Difference (IS1-IS2)	& $<2.5 \times{}10^{-11}$ $\pm{}0.5\times{}10^{-11}$ at 1$\sigma$ \\
SUEP Difference (IS1-IS2) & $< 11.6 \times{}10^{-11}$ $\pm{}0.9\times{}10^{-11}$ at 1$\sigma$ \\
\br
\end{tabular}
\end{indented}
\end{table}

\subsubsection{Stiffness}

Most of the measurement offset comes from the stiffness of each sensor with respect to the instrument frame. 
The periodic displacement of a test mass with respect to its electrodes induces a measurable periodic acceleration proportional to its stiffness and to its displacement. The sensor stiffnesses have been characterised in flight during the commissioning phase (Table \ref{tab_stiffness}).

\begin{table}[h]
\caption{\label{tab_stiffness} Measured and expected (between brackets) stiffness; the theoretical values have been computed before the flight assuming a perfect and simple electrostatic configuration and a negligible stiffness of the wire.}
\begin{indented}
\item[]\begin{tabular}{@{}lllll}
\br
Axis & IS1-SUREF & IS2-SUREF & IS1-SUEP & IS2-SUEP \\
\mr
$X$ [$\times{}10^{-3}\rm N\,m^{-1}$]& 1.1 ($\sim 0$) & 5.7 (0) &  1.8 ($\sim 0$) & 1.1 (0) \\
$Y$  [$\times{}10^{-2}\rm N\,m^{-1}$] & -1.5 (-2.8) & -8.4 (-14.3) & -1.7 (-2.8) & -6.9 (-12.4) \\
$Z$  [$\times{}10^{-2}\rm N\,m^{-1}$] & -1.5 (-2.8) & -7.5 (-14.3) & -1.5 (-2.8) & -6.8 (-12.4) \\
$\Phi$ [$\times{}10^{-5}\rm N\,rad^{-1}$]  & 3.9 (-0.8) & 345 (-0.7) & 1.2 (-0.8) & 6.5 (-0.7) \\
$\Theta$ [$\times{}10^{-2}\rm N\,rad^{-1}$] & -0.6 (-0.7) & -3.4 (-5.3) & -0.5 (-0.7) &  -3.0 (-3.5) \\
$\Psi$ [$\times{}10^{-2}\rm N\,rad^{-1}$]  & -0.6 (-0.7) & -3.9 (-5.3) & -0.6 (-0.7) & -3.0 (-3.5) \\
\br
\end{tabular}
\end{indented}
\end{table}

The stiffness is expected to be particularly low on the $X$-axis because the capacitive sensing depends on the variation of overlap of the electrode area on the test-mass \cite{hudson07}. It should be similar on $\Phi$. But as shown in Table \ref{tab_stiffness}, both $X$ and $\Phi$ have a significant positive stiffness due to the higher than expected gold wire stiffness: a result of modifying the integration process to improve the resistance to launch vibrations.
On the contrary, along the $Y$ and $Z$ axes, and for $\Theta$ and $\Psi$ rotations, the capacitive sensing is based on the variation of gap which generates a negative stiffness as for $\Theta$ and $\Psi$ sensing. The electrostatic stiffness, the derivative of the force with respect to the displacement, depends only on the geometry and the voltage applied on the electrodes. That is why we observe similar values on $Y$ and $Z$, for the SUEP and SUREF internal mass, that are electrostatically identical. For the external masses, the difference of the electrostatic stiffness is due to the different voltages applied on electrodes. The differences between the model and the in-orbit estimation is due to the simplified theoretical electrostatic configuration. The sensitivity of stiffness to the square of the voltage has been checked on all axes (but $X$) during the assessment phases, proving mainly an electrostatic origin; this has been checked by comparing stiffness when operating the sensors either in FRM (with $V_p = 40$V) with that in HRM (with $V_p = 5$V): stiffness varies with $V_p^2$. 
Note that for the axes $Y$, $Z$, $\Theta$ and $\Psi$, the major contribution to the stiffness comes from the outer cylinder supporting the $X$ and $\Phi$ electrodes. Thus, a larger gold wire stiffness causing a higher offset along $X$, the PID controller applies a DC voltage on $X$ electrodes increasing the stiffness along the radial axes. That is why the measured stiffness value for IS2-SUEP is higher that expected.

The observed orders of magnitude confirm the accuracy of the geometry but the stiffness along the $X$-axis is larger than expected and independent of $V_p$, most likely because of the gold wire and its implementation \cite{willemenot00b,willemenot00a}. This error source is independent of the electrode geometry but depends on the geometry of the wire when glued by its extremities: the tools used to handle such a thin wire do not allow for a full control of its initial geometry; flexure and traction can mix when the mass moves, leading to a large range  of values.

The stiffness, either negative or positive, leads to an offset in the restoring force. As the mechanical stiffness of the gold wire was higher than expected, it was decided to increase the range of measurement along $X$ by applying a DC voltage on the $X$ electrodes $V'_p = -2.5$V.  This voltage changes the scale factor of the $X$-axis and thus the sensor dynamic response.

\subsection{Thermal sensitivity} \label{ssect_therm}

From a thermal point of view, the instrument is composed of three elements: the digital electronics in the ICU, the two FEEU analogue electronics units and the two SU housings including the two masses. Each one has its own temperature, whose variation impacts the measurement in different ways:
\begin{itemize}
\item the variation of the electronics temperature induces a variation of the reference voltages, leading to a variation of scale factor or acceleration offset;
\item the variation of the SU temperature induces a variation of the instrument's geometry and a variation of the force offsets due to photon (radiation pressure), to outgassing and to residual gas pressure (radiometer effect). The last two depend on the anisotropy of the temperature stability and the residual gas pressure. They are negligible primarily because the temperature stability is much better than the specified 0.001K at $f_{EP}$ (the anisotropy has been also measured and is 3 times lower). The pressure specified to be lower than $10^{-5}$Pa has not been measured in flight but was measured to be within the requirements on an engineering model during more than 5 years in the laboratory.
\end{itemize}

The temperature of the ICU is additionally constrained by the stability need of the bus power converters placed near the mechanical interface of the satellite. However, these specifications are compatible with the digital electronics operation and have no impact on the performance.

Despite the very small temperature variations of the FEEU and of the SU around $f_{\rm EP}$, we do observe a drift of the acceleration measurements at lower frequencies; it is due to a temperature sensitivity. However, this can be corrected (see Sect. \ref{sect_analysis}), so that its impact remains negligible at $f_{\rm EP}$. We noted in Ref. \cite{touboul17} that the acceleration sensitivity to SU temperature variations is two orders of magnitude higher than expected. 
Since then, additional measurements allowed us to explain it as a consequence of thermal expansion of the satellite interface at SU bindings: SU parts expand more than the expected values calculated only with SU material properties. This expansion causes elongation of the gold wire of the outer test-mass and thus an increased force.

\begin{figure}
\center
\includegraphics[width=0.7\textwidth]{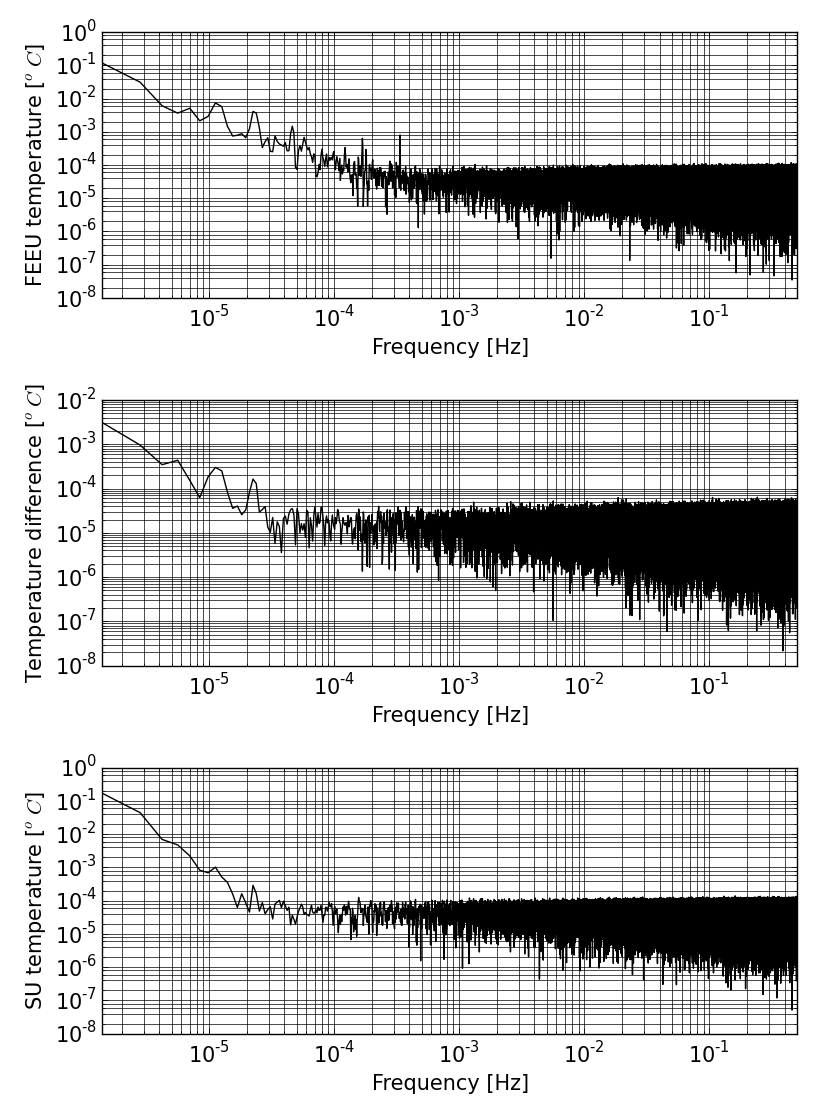}
\caption{SUEP temperature variations: FEEU (upper panel), difference of temperature inside SU between 2 probes separated by 159 mm along $X$ (middle panel) and SU (lower panel).}
\label{fig_thermal}       
\end{figure}

The temperature of the FEEU is measured by 5 Pt-resistance thermometers mounted on the circuit boards and the unit interface. The spectrum of the temperature measured by the probe located at the electronic interface ($T_{\rm FEEU}$) is shown in Fig. \ref{fig_thermal}.  At the $f_{\rm EP}$ frequency, no signal emerges from the probe noise of $2\times{}10^{-2}$\,K\,Hz$^{-1/2}$. That leads to a 1$\sigma$ upper bound $20 \times{}10^{-6}$ K of temperature variation at $f_{\rm EP}$. The requirement for the capacitive position sensor implemented in the FEEU is  a stability of 1 K Hz$^{-1/2}$ (equivalently, $2 \times{}10^{-4}$ K over 120 orbits), the thermal behaviour at the orbital frequency defining the worst case for the satellite design. As expected, spinning the satellite significantly improves (by up to two orders of magnitude) the temperature stability about $f_{\rm EP}$, mainly because of thermal filtering.

Similarly, the temperature of the SU has been measured in orbit at six locations to confirm the very good passive insulation of the satellite payload enclosure and the very low thermal dissipation inside the core of the sensors. The temperature fluctuations at the interface with the satellite,  $T_{\rm SU}$, are evaluated with the two probes closest to the interface. The temperature measurement is limited by the temperature probe noise, providing a 1$\sigma$ upper bound $15 \times{}10^{-6}$ K at $f_{\rm EP}$.

We looked for systematic error at $f_{\rm EP}$ due to  thermal variations with dedicated experiments. We could estimate the SU's and FEEU's thermal sensitivity by varying the temperature at the SU and FEEU interfaces with a controlled profile.

\begin{figure}
\center
\includegraphics[width=0.7\textwidth]{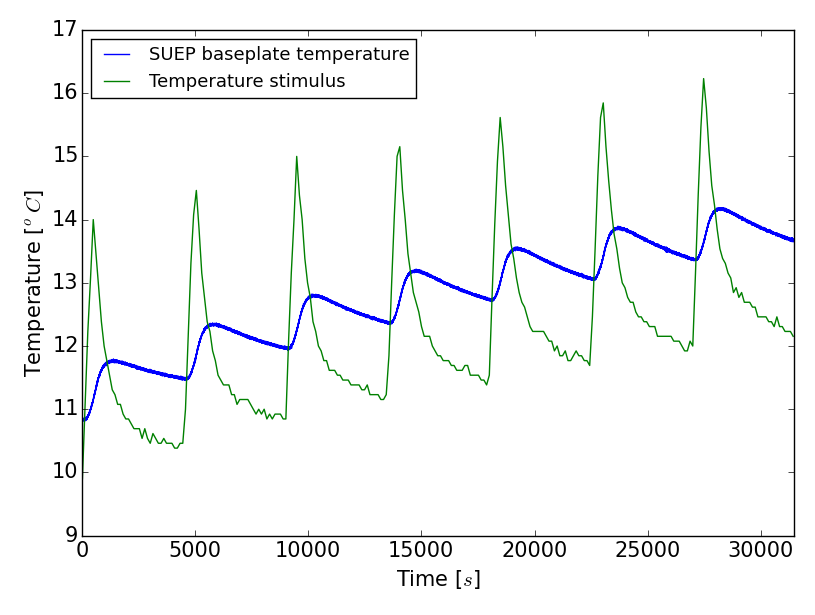}
\caption{Profile of the temperature stimulus (switch on of a resistor during 200sec periodically every 4500sec --green) and resulting temperature variations at the SUEP baseplate interface (blue).}
\label{fig_thermalstimulus}       
\end{figure}

Fig. \ref{fig_thermalstimulus} illustrates the experimental procedure: a temperature stimulus is locally applied to one of the units, FEEU or SU; in this particular case the resistors located on the plate between the two SU are switched on and off periodically in order to generate a periodic variation of temperature. The resistors are mounted by pairs in such a way that the current passing through each resistor of the pair is opposite and thus the induced magnetic field can be cancelled. Several periods have been used during the mission. The green line of Fig. \ref{fig_thermalstimulus} shows the temperature stimulus, while the blue line shows the concomitant SUEP baseplate temperature. The temperature and acceleration measurements are then analysed at the frequency of the stimuli, $f_{\rm th}$ and at its harmonics $2f_{\rm th}$, $3f_{\rm th}$ and $4f_{\rm th}$. The sensitivity at $f_{\rm EP}$ is deduced by interpolating the results from these 4 frequencies. Table \ref{tab_Tsens} lists the results for the $X$-axis.

\begin{table}[t]
\caption{\label{tab_Tsens} Difference of acceleration thermal sensitivity at $f_{\rm EP}$}
\begin{indented}
\item[]\begin{tabular}{@{}lcc}
\br
& SUREF & SUEP \\
\mr
Sensitivity to $T_{\rm SU}$ at $f_{\rm EP}$ [ms$^{-2}$K$^{-1}$] & $3.9\times{}10^{-9}$ & $4.3\times{}10^{-9}$ \\
Sensitivity to  $T_{\rm FEEU}$, at $f_{\rm EP}$ [ms$^{-2}$K$^{-1}$] & $5\times{}10^{-11}$ & $7\times{}10^{-11}$ \\
\br
\end{tabular}
\end{indented}
\end{table}

Two important remarks must be raised:
\begin{itemize}
\item The temperature variation stimuli are performed at the SU interface. These stimuli also generate a temperature gradient, measured with two probes in the SU separated by 159 mm along $X$. During the scientific sessions, the sources of the temperature variations at $f_{\rm EP}$ are located outside of the payload enclosure,  which is thermally decoupled from the rest of the satellite. Thus, the temperature variations around the SU are more uniform and the thermal sensitivity is dominated by the sensitivity to the interface temperature variation.
\item At the time of writing, no signal at $f_{\rm EP}$ has been detected in the temperature probe measurements during the science sessions. The values taken into account here are limited by the noise of the measurement pick-up considered at $1\sigma$.
\end{itemize}

\subsection{Disturbing field environment} \label{ssect_field}

\subsubsection{Magnetic field environment}

Because of their different magnetic susceptibilities, the  Pt and Ti test masses have different magnetic behaviours.
Consequently, the instruments are inside a magnetic shield whose efficiency was characterised on ground prior to the launch. The satellite's magnetic torquers are switched off (except during the commissioning phase) in order to minimise any magnetic source on board. In the same way, particular care was taken when designing the sensor to avoid electrical pins with magnetic moments. The magnetic moment of the satellite is lower than 0.2 A\,m$^2$ (the specification was 1 A\,m$^2$) as deduced from the level of the residual torque due to the Earth's magnetic field, as it is counteracted by the satellite attitude control. 
Finally, a 3D finite element model of the satellite and of the instrument was realised to assess the residual magnetic field and gradient at the test-mass level. The magnetic field variations and its effects in terms of acceleration at $f_{\rm EP}$ were also considered.

\subsubsection{Electric field environment}

The vacuum tight metallic housings of the instruments act also in orbit as an electrical shields for both sensors. We have not observed any disturbance in the feedthroughs (that may limit the shielding). Measurements of the electronics were performed in open loop during the instrument and satellite Electro-Magnetic Compatibility (EMC) acceptance tests. In particular we compared the noise of the capacitive sensor at each stage with the reference one obtained with the electronics connected to a reference capacitor in the laboratory. No disturbing signal was detected, proving the low effect of the feedthroughs.

Since the test-mass voltage is controlled by the gold wire, we do not consider the test-mass electrical charge and its fluctuation due to particle radiation.

\subsubsection{Local gravity field environment}

Besides the Earth's gravity, one has to consider the gravity and gravity gradient due to the satellite itself.  
A detailed model with meshing based on a Computer-Aided Design (CAD) model of the satellite and payload has been computed to estimate the self-gravity at the test-mass position. This model allowed us to check the thermal expansion due to temperature variation at different frequencies. Indeed, the thermal expansion at $f_{\rm EP}$ makes the mass distribution move and thus generates local gravity field variations.
It has thus been demonstrated that their effects are negligible, even when we consider the motion of the test-mass inside the same instrument (note that the satellite has been designed to have no moving parts). The major effects come from the distribution of mass nearest to the test-mass (i.e. the SU itself). The geometry and the material used for the SU are well defined, so that the CAD model is well suited to compute the local gravity distribution and variations in the worst case conditions.

Table \ref{tab_grav}  summarises the local gravity effect in the differential measurement (maximum value along all axes) evaluated by considering the thermal variations as specified in inertial pointing. In spin mode, the temperature variations are much lower and should result in even smaller effects.

\begin{table}
\caption{\label{tab_grav} Gravity perturbations: satellite local gravity variations established by modelling}
\begin{indented}
\item[]\begin{tabular}{@{}ll}
\br
DC value of gravity field & $1.8\times{}10^{-8}$m\,s$^{-2}$ \\
Gravity field variations at $f_{\rm EP}$ (common mode effect) & $4\times{}10^{-13}$m\,s$^{-2}$ \\
Gravity gradient at $f_{\rm EP}$ & $4\times{}10^{-13}$ s$^{-2}$ \\ 
{ }{ }{ }This effect results in $8\times{}10^{-18}$m\,s$^{-2}$ disturbing difference of acceleration\\
{ }{ }{ }with 20$\mu$m off-centring \\
Difference of acceleration due to gravity effect & $1.5\times{}10^{-17}$m\,s$^{-2}$ \\
 (thermal expansion) and  to test-mass shape defects \\
\br
\end{tabular}
\end{indented}
\end{table}

\subsection{Summary of systematic error sources}

Table \ref{tab_calibration} summarises the distribution of systematic error sources and the method used in the analysis to evaluate their amplitude or upper bound. The effect of off-centrings are evaluated by in orbit calibration associated to the DFACS performances. The DFACS performances are established with the accelerometer common mode and the star-tracker measurements (see Sect. \ref{ssect_dragfree}). \\
The main source of error comes from the temperature variation at  $f_{\rm EP}$ seen through the accelerometer sensitivity (Sect. \ref{ssect_therm}). The value used to establish the systematic error is calculated on the basis of the noise of the temperature probe integrated over 120 orbits at 1$\sigma$ for the variations at $f_{\rm EP}$. Recent analyses, still under validation, may show that the actual variation is much lower. Nevertheless, in this paper we remain conservative and keep the more recent analyses for an upcoming paper.

For the sake of simplicity, and to avoid a detailed consideration of correlation between errors, we add systematics linearly to get a 1$\sigma$ upper bound from systematics expressed as an E\"otv\"os parameter of $9\times 10^{-15}$.

Table \ref{tab_calibration} shows the result for the SUEP. The error allocation for SUREF is the same except for thermal systematics which are estimated as $61 \times{}10^{-15}$\,m\,s$^{-2}$ if we consider the thermal sensitivities of Table \ref{tab_Tsens}. Thus, the total systematic error for SUREF, expressed as an E\"otv\"os  parameter is $8 \times{}10^{-15}$ at 1$\sigma$.

\begin{table}
\caption{\label{tab_calibration} Evaluation of systematic errors in the difference of acceleration measurement for SUEP @$f_{\mathrm{EP}}$=3.1113$\times{}10^{-3}\,Hz$.}
\begin{indented}
\item[]\begin{tabular}{@{}lll}
\br
{\bf Term in the Eq.\,(1) projected}  & \vline {\bf  Amplitude or } & \vline {\bf Method} \\
{\bf on $\overrightarrow x$ in phase with $g_x$ at $f_{\mathrm{EP}}$} & \vline {\bf upper bound} & \vline {\bf of estimation} \\ \hline
{\bf Gravity gradient effect } &  & \\ 
{\bf $\left[ T \right]\overrightarrow \Delta$ in \,m\,s$^{-2}$} &  & \\ \hline
($T_{xx}\Delta x$; $T_{xy}\Delta y$; $T_{xz}\Delta z$)  & \vline $< (10^{-18}$;$ 10^{-19}$;$ 10^{-17}$)  & \vline Earth{'}s gravity model. \\ \hline
{\bf Gradient of inertia matrix $\left[ In \right]$} &  &  \\
{\bf effect along $X$ in\,m\,s$^{-2}$ } &  &  \\ \hline
 & \vline & \vline DFACS performances \\
$\dot \Omega_y\Delta z - \dot \Omega_z \Delta y$ & \vline $5\times{}10^{-17}$ & \vline and calibration. \\  \hline
$\Omega_x \Omega_y \Delta y - \Omega_x \Omega_z \Delta z$   & \vline  & \vline DFACS performances \\ 
$- \left(\Omega_y^2 + \Omega_z^2 \right)\Delta x$ & \vline $1.3 \times{}10^{-17}$ &  \vline and calibration. \\ \hline
{\bf Drag-free control in \,m\,s$^{-2}$} & {} & \\ \hline
 & \vline & \vline DFACS performances \\
$(\left[ M_d \right] \overrightarrow \Gamma_c^{app}). \overrightarrow x$ & \vline $1.7 \times{}10^{-15}$ & \vline and calibration. \\ \hline
{\bf Instrument systematics} & & \\ 
{\bf and defects in \,m\,s$^{-2}$} & & \\ \hline
 & \vline & \vline DFACS performances \\
$(\overrightarrow \Gamma_d^{quad}).\overrightarrow x$  & \vline $5 \times{}10^{-17}$ & \vline and calibration. \\ \hline
$([Coupl_d] \dot {\overrightarrow \Omega}).\overrightarrow x$ & \vline & \vline Couplings observed \\ 
 & \vline $< 2 \times{}10^{-15}$ & \vline during commissioning phase.\\ \hline
Thermal systematics & \vline & \vline Thermal sensitivity \\ 
 & \vline $<67 \times{}10^{-15}$ & \vline in-orbit evaluation.  \\ \hline
Magnetic systematics & \vline $< 2.5 \times{}10^{-16}$ & \vline Finite elements calculation.\\ \hline
{\bf Total of systematics in $\Gamma_{dx}^{meas}$} & \vline {\bf $< 71 \times{}10^{-15}\,$m\,s$^{-2}$}  & \\ \hline
{\bf Total of systematics in $\delta$ } & \vline $< 9 \times{}10^{-15}$ & \\
\br
\end{tabular}
\end{indented}
\end{table}

\subsection{Signal stationarity by wavelet analysis}

So far, we based our analyses on power spectra, i.e. on FFT of the autocorrelation of the measured accelerations, under the implicit assumption that the signal is stationary. However, the Fourier transform does not provide any temporal information and is clearly not suited to detect non-stationarities in the data (either transients or slow drifts). A non-stationarity in the data will plague our analysis and hamper our estimation of the E\"otv\"os parameter.

\begin{figure}
\includegraphics[width=0.95\textwidth]{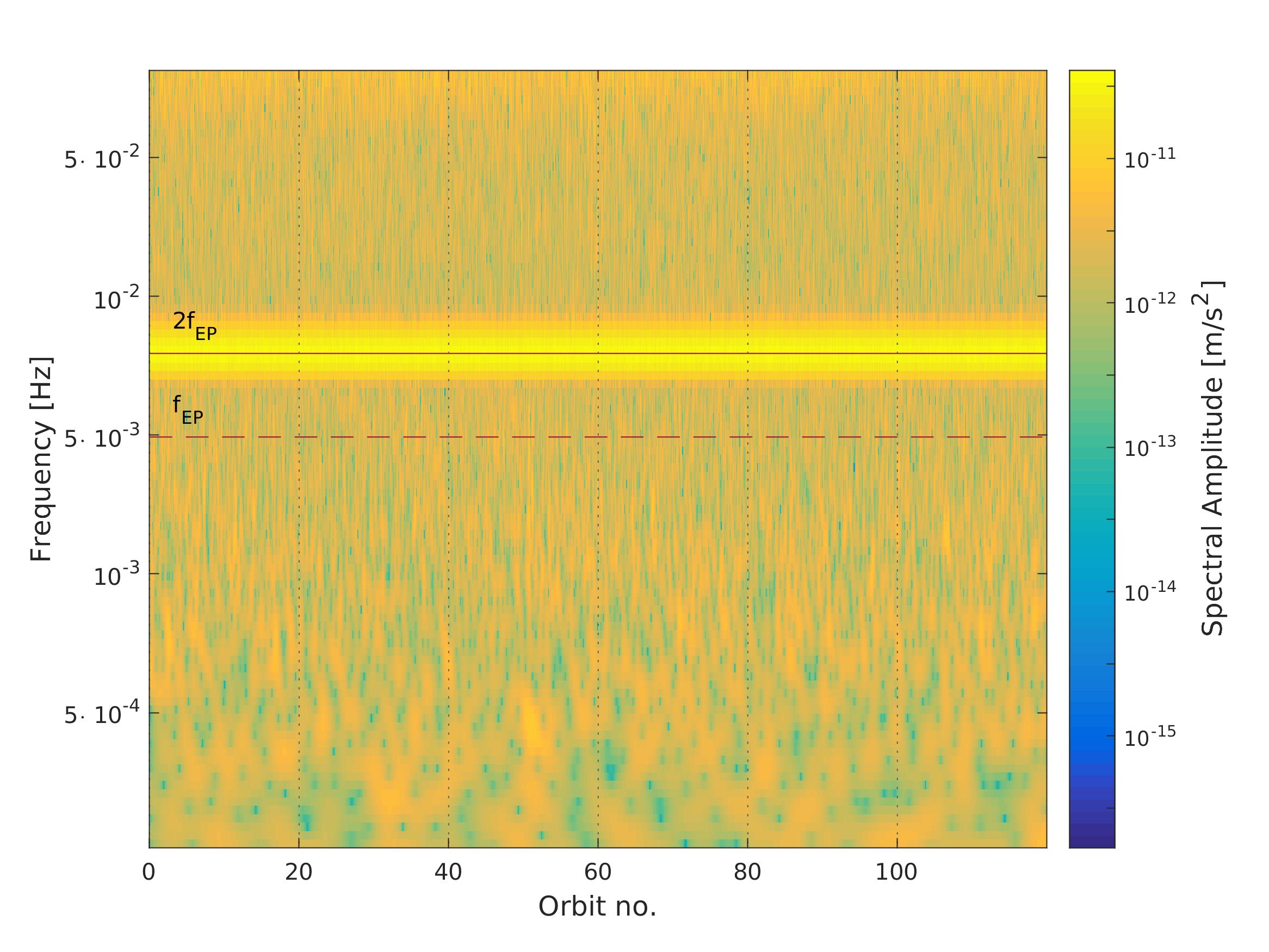}
\caption{Wavelet representation of SUEP difference of acceleration measurement. The solid horizontal line indicates the characteristic frequency $2 f_{\rm EP}$, the dashed line $f_{\rm EP}$. The colour indicates the fraction of the total signal energy.}
\label{fig_wavelets}       
\end{figure}

In this paper, we checked the stationarity of the data using a wavelet analysis \cite{selig16} that provides a time-frequency representation of the signal. Fig. \ref{fig_wavelets} shows the obvious off-centring signal at $2 f_{\rm EP}$ (solid line), but  no significant continuous or temporal signals at $f_{\rm EP}$ (dashed line); in particular, no frequency-varying signal is detected.

In future analyses, the stationarity of the noise and/or signals will be assessed with better sensitivity using wavelet analysis due to the accumulation of all the available sessions.

\section{Data analysis and results} \label{sect_analysis}

When analysing EP sessions, we estimate three parameters simultaneously: the approximated E\"otv\"os parameter $\delta$ and the components $\Delta x$ and $\Delta z$ of the off-centring. We use the model equation (\ref{eq_gammad}) and proceed in several steps:
\begin{enumerate}
\item we first fit the measurements $\overrightarrow\Gamma_d^{meas}$ with a polynomial of order 3, in particular to correct the effects of long term drift with the temperature;
\item we then correct the measurements $\overrightarrow\Gamma_d^{meas}$ with the in-flight calibrated $[M_d]$ matrix and with the off-centring along all axes also estimated with the calibration sessions associated to the estimation of the Earth's gravity gradient. Quadratic terms are also estimated but do not need to be corrected as their effect is found to be negligible; 
\item then we can use the simplified equation $\overrightarrow\Gamma_d^{meas}(t_i) =\delta g_x(t_i) + T_{xx}(t_i) \Delta x + T_{xz}(t_i) \Delta z$ for the $N$ dates of the measurement $t_i \,\, (0 \leqslant t_i \leqslant T)$ sampled at $t_s =0.25 \,s$;
\item these equations are projected in the frequency space by applying a discrete Fourier transform to the time series $\overrightarrow\Gamma_d^{meas}(t_i)$, $g_x(t_i)$, $T_{xx}(t_i)$ and $T_{xz}(t_i)$ to get $\overrightarrow\Gamma_d^{meas}(f_j) =\delta g_x(f_j) + T_{xx}(f_j) \Delta x + T_{xz}(f_j) \Delta z$ for the $N/2$ frequencies $f_j \,\,  (0\leqslant f_j \leqslant 1/(2 t_s))$ with a sampling of $f_T=1/T$; for each frequency we get complex values including a real and an imaginary part;
\item since the useful signal is concentrated at $f_{\rm EP}$ for the E\"otv\"os parameter and at $2 f_{\rm EP}$ for the off-centring, we select narrow bands around these frequencies: this is equivalent to selecting the corresponding equations in the frequency domain;
\item from these selected equations, the parameters $\delta$, $\Delta x$, and $\Delta z$ are estimated by weighted least-square method; the weighting is a diagonal matrix using the inverse of the estimated measured Power Spectral Density (PSD) for each frequency.
\end{enumerate}

During measurement sessions, gaps in the data may occur because of many reasons such as data losses of few seconds in the instrument (less than two per year), very short (one data point) losses between the satellite memory system and the ground segments (at most one to two times per day). 
These gaps induce leakage phenomena in the frequency domain, leading to an apparent increase of the noise level at low frequency \cite{baghi15}. Several methods have been developed to overcome this effect, either by generalising the least-square regression technique \cite{baghi15, baghi16} or by filling the gaps with artificial data consistent with the statistics of the observed data \cite{berge15b, pires16}.

The unique gaps in these two cases come from telemetry data losses: only 8 points are missing in the two sessions analysed in this paper, which represents less than 0.001\% of the data. These few gaps have been corrected  with the inpainting method \cite{berge15b,pires16}, but we have verified that with so small a number of gaps (more than 1000 times smaller than the worst case anticipated before the launch), computing the missing values as a local average of the neighbouring data yields similar results. 

Our analysis, done using the two sessions described above, provides constraints on the E\"otv\"os parameter (i.e. acceleration divided by the amplitude $g_x=7.9\,m\,s^{-2}$) one order of magnitude better than pre-MICROSCOPE measurements \cite{wagner12}:
\begin{equation}
\delta({\rm Ti,Pt})=[-1\pm{}9{\rm (stat)}\pm{}9{\rm (syst)}] \times{}10^{-15} \quad (1\sigma \,\,  {\rm statistical \,\, uncertainty}),
\end{equation}
where the systematic error is dominated by thermal effects, as shown in Table \ref{tab_calibration}.

We used the same analysis process to extract a signal at $f_{\rm EP}$ from the SUREF difference of acceleration measurement, leading to:  
\begin {equation}
\delta({\rm Pt,Pt})=[+4\pm{}4{\rm (stat)}\pm{}8{\rm (syst)}] \times{}10^{-15} \quad (1\sigma \,\,  {\rm statistical \,\, uncertainty}),  
\end {equation}
which is compatible with a null value as expected.

This result was obtained from raw measurements. The calibration sessions (scale factors matching and misalignment estimations) were used only to validate the good behaviour of the system and to confirm the requirements on the matrix $[M_d]$. In the forthcoming analyses aimed to reach the mission objective of $10^{-15}$, the calibration of matrix $[M_d]$ may be necessary.
The error quoted in the estimation of the E\"otv\"os parameter is the root of the variance of the least-squares method in the Fourier domain, rescaled by the root mean square of the residuals.


\section{Conclusion and perspectives} \label{sect_ccl}

In this paper, we presented the first results of the MICROSCOPE mission. We provided details about the mission, the satellite, the instrument, our assessment of systematic errors, as well as our data analysis process, which allowed us to consolidate the results given in letter \cite{touboul17}.

In particular, the matching of the scale factors and alignments of the instrument were performed in orbit with a sensitivity better than specified. The very good performance of the satellite's drag-free system allowed for the relaxation of some constraints on the effect of the common mode accelerations. The data analysis did not show any evidence for the presence of a differential signal between platinum and titanium alloys at $f_{EP}$ and at 1$\sigma$ statistical uncertainty: $\delta({\rm Ti,Pt})=[-1\pm{}9{\rm (stat)}\pm{}9{\rm (syst)}] \times{}10^{-15}$.

This result takes into account the estimation of the systematic errors and the measured variances for the statistical error over 120 orbits . The systematic errors are dominated by thermal effects, which will be further analysed and better estimated in a future work. Most importantly, albeit this preliminary conclusion seems robust, it has to be re-assessed after cumulating all the 1800 orbits for SUEP and the 980 orbits for SUREF that are available today.

The MICROSCOPE in orbit mission came to its end in October 2018. Additional scientific data are under validation and should improve the current result as the final amount of data represents about 15 times the amount analysed in this paper.


\ack

The authors express their gratitude to all the different services involved in the mission partners and in particular CNES, the French space agency in charge of the satellite. This work is based on observations made with the T-SAGE instrument, installed on the CNES-ESA-ONERA-CNRS-OCA-DLR-ZARM MICROSCOPE mission. ONERA authors' work is financially supported by CNES and ONERA fundings.
Authors from OCA, Observatoire de la C\^ote d'Azur, have been supported by OCA, CNRS, the French National Center for Scientific Research, and CNES. ZARM authors' work is supported by the German Space Agency of DLR with funds of the BMWi (FKZ 50 OY 1305) and by the Deutsche Forschungsgemeinschaft DFG (LA 905/12-1). The authors would like to thank the Physikalisch-Technische Bundesanstalt institute in Braunschweig, Germany, for their contribution to the development of the test-masses with funds of CNES and DLR.

\section*{References}
\bibliographystyle{iopart-num}
\bibliography{miccqg}

\end{document}